\documentclass[showpacs, preprintnumbers, amsmath, amssymb, twocolumn]{revtex4}

\usepackage{graphicx}
\usepackage{dcolumn}
\usepackage{bm}
\usepackage{subfigure}

\begin{document}

\preprint{...}

\title{Equation of motion for density distribution of many circling particles with an overdamped circle center}

\author{Tieyan Si}
\affiliation{Max-Planck-Institute for the Physics of Complex Systems. Nothnitzer Street 38, D-01187 Dresden,
Germany}

\date{\today}

\begin{abstract}

We first established the dynamic equations to describe the noisy circling motion of a single particle and the 
corresponding probability conservation equation in both two dimensions and three dimensions, and then 
developed the evolution equation of density distribution of many circling particles with overdamped circle center. 
For many circling particle system without any external force, the density gradient in one direction can 
induce a flow perpendicular to this direction. While for single circling particle, 
similar phenomena occurs only for non-zero external force. We performed numerical 
evolution of the density distribution of many circling particles, the density distribution 
behaves as a decaying Gaussian distribution propagating along the channel. 
We computed the particle flow field and the effective force field. 
Vortex shows up in the high density region. The force field drive particles 
to the transverse direction perpendicular to the density gradient.
 We applied this non-equilibrium evolution equation to understand the diffusion phenomena
 of many sperms(\textit{J. Exp. Biol.} \textit{210}, \textit{3805-3820}). Numerical evolution gave us similar
 density distribution as experimental measurement. The transverse flow we predicted provide 
a theoretical understanding to the bias concentration of many sperms(\textit{J. Exp. Biol.} \textit{210}. \textit{3805-3820}).

\end{abstract}


\maketitle

\tableofcontents

\section{Introduction}

Particles moving along a circular trajectory appears everywhere in nature. The most familiar examples are the moon and electron. 
Living microorganism moving along circles is unusual. The bacterium $Escherichia \;coli$ swim in circles near 
a surface\cite{lauga}\cite{diluzio}\cite{hill}\cite{berg}. Sea urchins spermatozoa swim along a helical trajectory in ocean. 
When they reach the surface of ocean, they swim along a noisy circle\cite{riedel}\cite{woolley}\cite{Bohmer}\cite{eisenbach}. Some 
artificial particles also demonstrate circular motion, such as red blood cell attached by a chain of colloidal magnetic 
particles\cite{dreyfus} and double faced nanorodes\cite{walther}. Living microorganisms swim in thick liquid 
with low Reynolds number. Hydrodynamics at low Reynolds number may play an important role\cite{purcell}\cite{pedley}\cite{dean}\cite{cates}\cite{marchetti}.   
Recently, the circular motion of living microorganism has attracted a lot of theoretical interests which covered 
various different biological mechanism\cite{lauga}\cite{diluzio}\cite{hill}\cite{berg}\cite{camalet}
\cite{dhar}\cite{teeffelen}\cite{julicherPNAS}\cite{grenshaw}\cite{julicherNJ}\cite{yang}\cite{li}.

So far, there is no analytic theory to study the general physics of circling particle with overdamped circle center. 
In order to demonstrate the common physics of various different circle swimmers, 
we study a type of abstract point particle as analogy of those tiny biological microorganism. This abstract point particle 
inherit the main characteristic motion 
of those circling microorganism. They bear the following characters: 
(1) the point particle moves along a circle. The center of the circle is drifting around. 
(2) The circling speed is much larger than the drifting speed of the center so that one can observe 
an apparent circular trajectory. (3) The particle is moving in a highly viscous environment which 
represents the thick liquid filled with various different molecules, proteins and other microorganism. (4) 
The drifting motion of the circle center is overdamped, because the slow drifting of the circle center 
is too weak to counterbalance frictional resistance. The local circling motion around the 
circle center is not overdamped, because the self-propel force canceled  
the frictional force in the tangent direction along the circle. (5) The particle obeys Newton's law of 
motion. The force in Newton's law of motion is the superposition of deterministic force and stochastic force. 
We put the stochastic force and deterministic force together to get unified theoretical description.

The abstract circling particle is a theoretical mapping of circling bacteria and 
sperm. Both the bacterium $Escherichia\; coli$ and sea urchins spermatozoa are macroscopic 
objects comparing to the scale of water molecules. When there is no any other external 
chemical source, the circle center of a single circling bacteria or a sperm randomly drift around\cite{woolley}. 
In the presence of chemoattractant, the center began drifting along some 
bias direction\cite{riedel}\cite{woolley}. 
In both cases, the trajectory is well approximated by drifting circles. 
An apparent circular trajectory requires the local circling speed should be much higher than the drifting speed of 
the center. If the circle center moves faster than the local circling particle 
itself, the trajectory would not be drifting circles.

\begin{figure}[htbp]
\centering
\par
\begin{center}
\includegraphics[width=0.43\textwidth]{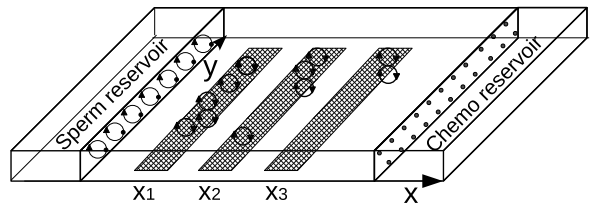}
\end{center}
\caption{\label{expe}The experiment device for the diffusion of
many sperms under attraction of chemoattractant in Ref.\cite{inamdar}. Each circle represents one sperm. The arrow labels the circling direction. The length of the channel is 7000-9000 $\mu{m}$, the width is 500 $\mu$m, and the height is 100 $\mu{m}$.} \vspace{-0.2cm}
\end{figure}

Recently an experiment had been conducted to study the diffusion phenomena of many \emph{Arbacia punctulata}
spermatozoa in seawater, jelly coat solution and resact solution\cite{inamdar}. 
The experiment observation in Ref.\cite{inamdar} is performed in a rectangular migration channel with
 sperm reservoir and chemoattractant attached to its two
ends(Fig. \ref{expe}). The diffusion image shows the sperm liken to concentrate one side of the channel. 
Sperms move faster near the boundary of the channel\cite{inamdar}. One motivation of my theoretical 
research is to understand the diffusion phenomena observed in Ref. \cite{inamdar}.

The organization of the article is following:

In section II, we derived the probability conservation equation of single circling particle both in 
two dimensions and three dimensions. We established the dynamic equation of single circling particle from 
Newton's law of motion. The force term is the sum of deterministic force and stochastic force.  
If there is no external forces, the center of single circling particle would randomly drift around. 
In the presence of an effective external field, such as electric field for charged particle or 
chemoattractant for bacteria and sperm, there exists a 
drifting velocity perpendicular to the direction of the external field.

In section III, we derived the evolution equation for the density distribution of many circling particles 
from Newton's law of motion and the conservation equation of total particle number. 
The general Fokker-Plank equations is a special approximation of the 
evolution equations we derived.

In section IV, we performed numerical computation of this evolution equation of density distribution. 
It shows particle number conservation equation captured most important physics 
of many circling particles system, and gave similar density distribution as experimental measurement. A drifting 
flow perpendicular to the density gradient shows up for the case without any external field. 
If the external field is too strong, the bias density concentration is suppressed. We also found   
vortex and turbulence configuration.

In section V, we applied the evolution equation of density distribution to understand the 
many sperm diffusion experiment\cite{inamdar}. We assumed that the action of chemoattractant on 
sperm is equivalent to an effective attractive force. Numerical evolution of the many particle 
diffusion equation shows a transverse flow perpendicular to the density gradient. 
The direction of this transverse flow depends on the direction of angular velocity. 
Similar density distribution as that of the experiment data in Ref.\cite{inamdar} is obtained.

Section VI is devoted to a brief summary. The detail calculation for deriving my equations 
are presented in the appendix.

\section{The probability conservation equation of single circling particle}

\begin{figure}[htbp]
\centering
\par
\begin{center}
\includegraphics[width=0.38\textwidth]{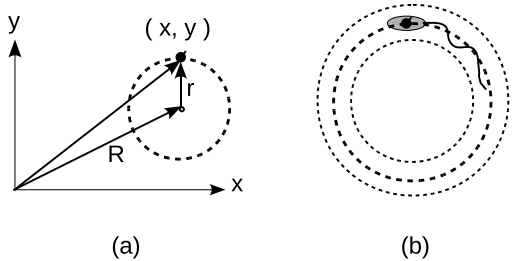}
\end{center}\vspace{-0.5cm}
\caption{\label{circle} (a) The coordinates of one circling particle. (b) Bacteria or sperm swims along a 
fluctuating trajectory which is confined in an annulus.} 
\end{figure}

The sperm or bacteria swim along a noisy circle (Fig. \ref{circle} (b)). We cover the noisy circle by a 
perfect circular tube. The sperm or bacteria is mapped to a point particle( Fig. \ref{circle} (a)). This point 
particle is moving along the center-line of the annulus( Fig. \ref{circle} (b)).

The vector denoting the instantaneous position of a circling particle may be decomposed as 
the sum of two vectors: one vector is the instantaneous center of the circle, 
$\vec{R}(t)=R_{x}(t)\textbf{e}_{x}+R_{y}(t)\textbf{e}_{y}$; the other vector measures 
the relative position with respect to the center, 
$\vec{r}(t)=r_{x}(t)\textbf{e}_{x}+r_{y}(t)\textbf{e}_{y}$. Vector $\vec{r}(t)$ is 
rapidly circling around the center at frequency $\omega$, 
\begin{eqnarray}
r_{x}(t)={r}_{0}(t)\cos({\omega} t),\;\;\;\;r_{y}(t)= {r}_{0}(t)\sin({\omega} t).
\end{eqnarray}
A general geometric description of a helical curve of drifting circle reads  
\begin{eqnarray}
x(t)\textbf{e}_{x}+y(t)\textbf{e}_{y}=\vec{R}(t)+\vec{r}(t). 
\end{eqnarray}
One may use differential geometry as another equivalent mathematical description of circling particles, 
however it is not convenient to study the motion of the drifting center and the local circling motion separately. 
The center has the freedom to move across the whole space, while the local circling motion is confined 
in a small circle with radius ${r}_{0}(t)$(one may choose ${r}_{0}(t)=constant$ for special cases, here we 
allow the radius to fluctuate). For a circling bacteria and sperm, the velocity of the drifting center 
is much slower than the local circling motion. The dominant dynamics of the circle center and the circling motion 
around the center is different.

A noisy circular trajectory is the output of three independent noisy forces: the external force on the circle center, 
the tangent force modifying the angular frequency and the fluctuating centripetal force. 
A stochastic trajectory itself can not tell which force results in the ultimate trajectory. 
The complete coordinates equation of a fluctuating circular trajectory is 
 \begin{eqnarray}
\tilde{x}(t)&=&[{R}^0_{x}(t)+{\xi}^R_{x}(t)]+[{r}_{0}(t)+{\xi}^r(t)]\cos[{\omega}^0+{\xi}^{\omega}(t)], \nonumber\\
\tilde{y}(t)&=&[{R}^0_{y}(t)+{\xi}^R_{y}(t)]+[{r}_{0}(t)+{\xi}^r(t)]\sin[{\omega}^0+{\xi}^{\omega}(t)]. \nonumber\\
\end{eqnarray}
The continuous function $({R}^0_{\alpha}(t), \;\;\alpha=x,\;y,)$ indicates the average position of the 
fluctuating part ${\xi}^R_{\alpha}(t)$ within a relatively longer time interval. 
The fluctuating radius is also denoted as a fluctuating factor upon the average length of the radius. 
The fluctuating part should be much smaller than the corresponding average length of the radius 
so that one is able to observe a fluctuating circle. Out of the same reason, we set the average frequency 
to a much larger value than the fluctuating term, i.e.,     
\begin{eqnarray}
\arrowvert{\xi}^r(t)\arrowvert\ll{r}_{0}(t),\;\;\;\;\;\;\arrowvert{\xi}^{\omega}(t)\arrowvert\ll\omega^0.\;\;
\end{eqnarray}

A noisy circle is the output of dominant deterministic motion over small fluctuations. 
If the deterministic force dominates the motion of a particle along a helical curve, 
we can define the tangent vector of the trajectory as velocity(see Appendix A). 
Acceleration can be related to the curvature of the trajectory. Circling motion is not inertial motion. The centripetal force only change
direction of the velocity while keep its absolute value invariant. When we say a particle is circling around a center, 
we have chosen a reference coordinates system in which a static observer stands far away from the particle and its center. 
If we put the reference coordinates system on the particle itself, 
an observer sitting on the particle would find the center of the original circle is circling around the particle. 
In fact, according to Newton's law of force, the center pulls the particle through centripetal force as well as 
the particle pulls the center through the same force in an opposite direction.

\begin{figure}[htbp]
\centering
\par
\begin{center}
\includegraphics[width=0.35\textwidth]{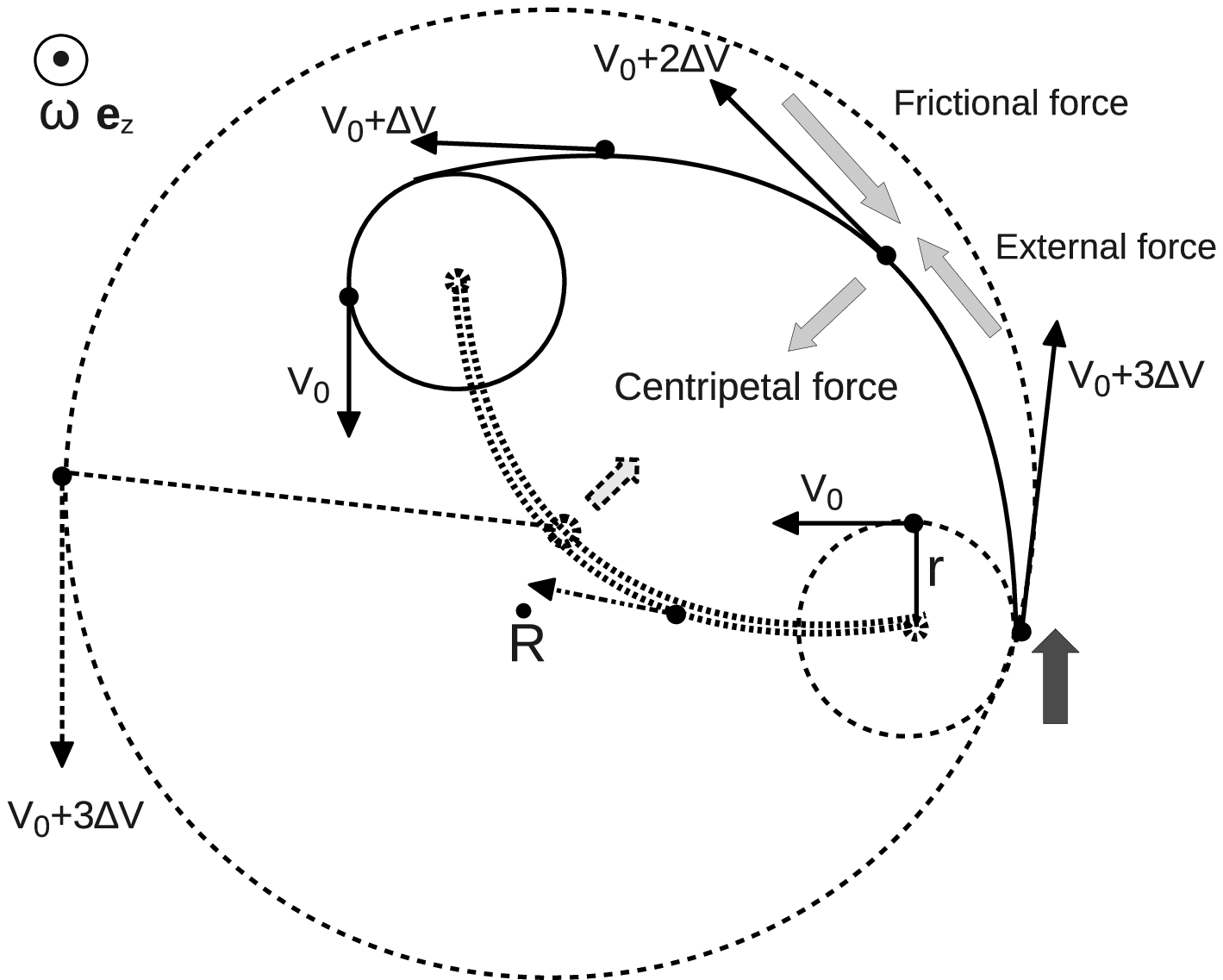}
\end{center}
\caption{\label{centri}. A schematic demonstration of the centripetal force upon the circle center. The double dashed-line is the 
trajectory of the center. The angular velocity points out of the paper. $V_0$ is the average velocity maintained by the self-propel force.} \vspace{-0.2cm}
\end{figure}

In the laboratory reference coordinates, the centripetal force exerted upon the circle center
by the local circling motion is expressed by angular frequency(see Appendix A for detail calculation), 
\begin{eqnarray}
F_{cen}=-{\dot{\vec{R}}}\times{\vec{\omega}}=-[\dot{\vec{R}}^0(t)+\dot{\vec{\xi}}^R(t)]\times
[{\vec{\omega}}^0+{\vec{\xi}}^{\omega}(t)].
\end{eqnarray}
This centripetal force $F_{cen}$ is a mathematical conception defined on 
a helical curve(see Appendix A for detail calculation). 
Here we use some continuous function ${\vec{\xi}}^R(t)$ as one approximation of the 
local fluctuating amplitude function so that $\dot{\vec{\xi}}^R(t)$ has a good mathematical definition. 
In fact, this centripetal force comes from the geometry of the circular trajectory. Imagine a particle 
circling around a static center at velocity $V_0$, when an external force promoted its velocity to 
$V_0+3\Delta{V}$, the centrifugal force is increased which in turn lengthen the radius. 
The static center of the original small circle is pushed to the new position of the center of the larger circle. 
As the particle is moving against high frictional force, the velocity will decrease, so does the radius of 
the instantaneous circle(Fig. \ref{centri}). 
The motion of the instantaneous center is driven by the centripetal force $-{\dot{\vec{R}}}\times{\vec{\omega}}$(Fig. \ref{centri}).

We divide the force in Newton's equation of motion for the circle center into three parts: the first is the centripetal force; 
the second is frictional force, $F_{friction}\propto-m \eta{\dot{\vec{R}}}$; the third is external force $F$. The 
Newton's equation of motion for the circle center reads
\begin{eqnarray}\label{2dyn-single0}
m\ddot{\vec{R}}=-m{\dot{\vec{R}}}\times{\vec{\omega}}+\vec{F}-m\eta{\dot{\vec{R}}},
\end{eqnarray}
where $\eta$ indicates the friction coefficient. The center coordinates $\vec{R}$, external force $\vec{F}$ and angular frequency $\omega$ are denoted as the sum 
of a deterministic term and stochastic term, 
\begin{eqnarray}
{R}_{x}&=&[{R}^0_{x}(t)+{\xi}^R_{x}(t)],\;\;\;\;{R}_{y}=[{R}^0_{y}(t)+{\xi}^R_{y}(t)], \nonumber\\
F_{x}&=&[F^0_{x}(t)+{\xi}^F_{x}(t)],\;\;\;\;F_{y}=[{F}^0_{y}(t)+{\xi}^F_{y}(t)], \nonumber\\
{\omega}&=&{\omega}^0+{\xi}^{\omega}(t) .
\end{eqnarray}
The Newton's equation with stochastic variables is
\begin{widetext}
\begin{eqnarray}\label{2dyn-single}
m[\ddot{\vec{R}}^0(t)+\ddot{\vec{\xi}}^R(t)]
=-m[\dot{\vec{R}}^0(t)+\dot{\vec{\xi}}^R(t)]
\times[{\vec{\omega}}^0+{\vec{\xi}}^{\omega}(t)]+[{\vec{F}}^0+{\vec{\xi}}^{F}(t)]-m\eta[\dot{\vec{R}}^0(t)+\dot{\vec{\xi}}^R(t)],
\end{eqnarray}
\end{widetext}
We decompose the equation of motion (\ref{2dyn-single}) into a pair of deterministic equations and a pair of 
stochastic equations, 
\begin{eqnarray}\label{2dyn-singleR}
\ddot{R}^0_{x}&=&-{\omega}^0\dot{R}^0_{y}-\eta\dot{R}^0_{x}+\frac{{F}^0_{x}}{m},\nonumber\\
\ddot{R}^0_{y}&=&{\omega}^0\dot{R}^0_{x}-\eta\dot{R}^0_{y}+\frac{{F}^0_{y}}{m},
\end{eqnarray}
\begin{eqnarray}\label{2dyn-single-xi}
\ddot{\xi}^R_{x}&=&-{\xi}^{\omega}(t)[\dot{R}^0_{y}+\dot{\xi}^R_{y}]-{\omega}^0\dot{\xi}^R_{y}-\eta\dot{\xi}^R_{x}
+\frac{{\xi}^F_{x}(t)}{m},\nonumber\\
\ddot{\xi}^R_{y}&=&{\xi}^{\omega}(t)[\dot{R}^0_{x}+\dot{\xi}^R_{x}]
+{\omega}^0\dot{\xi}^R_{x}-\eta\dot{\xi}^R_{y}+\frac{{\xi}^F_{y}(t)}{m}.
\end{eqnarray}
We assume the particle moves in an environment with high frictional force. 
The motion of the center is overdamped due to the high frictional force. 
The inertial effect may be ignored, thus we set them to zero,
\begin{eqnarray}
\ddot{R}^0_{x}=0,\;\;\ddot{R}^0_{y}=0,\;\;\;\;\ddot{\xi}^{R}_{x}=0,\;\;\ddot{\xi}^{R}_{y}=0.
\end{eqnarray}
Noticing here it is the circle center's inertial term. The local circling motion of the particle itself is not overdamped. 
The self-driven force of the local circling motion canceled the frictional force along the circle. 
For bacteria or sperm, the local self-driven force comes from the beating flagella which has bias beating directions. The beating flagella 
push it to move along a circle against the frictional force in the direction 
of the tangent vector of the circle. In the meantime, the beating flagella provide 
an effective centripetal force perpendicular to tangent vector which leads to a circular trajectory.

The deterministic velocity for the overdamped case is determined by Newtonian equation (\ref{2dyn-singleR}) 
and the constrain $\ddot{\vec{R}}^0=0$,
\begin{eqnarray}\label{1dotR}
\dot{R}^0_{x}(t)&=&\sigma^0_{xx}{F^0_{x}}+\sigma^0_{xy}{F^0_{y}},\nonumber\\
\dot{R}^0_{y}(t)&=&\sigma^0_{yy}{F^0_{y}}+\sigma^0_{yx}{F^0_{x}},
\end{eqnarray}
where the drifting tensor $\sigma^0$ are
\begin{eqnarray}
\sigma^0_{xx}&=&\frac{{m}^{-1}\eta}{\eta^2+[{\omega}^0]^2},\;\;\;
\sigma^0_{xy}=-\frac{{m}^{-1}[{\omega}^0]}{\eta^2+[{\omega}^0]^2},\nonumber\\
\sigma^0_{yy}&=&\frac{{m}^{-1}\eta}{\eta^2+[{\omega}^0]^2},\;\;\;
\sigma^0_{yx}=\frac{{m}^{-1}[{\omega}^0]}{\eta^2+[{\omega}^0]^2}.
\end{eqnarray}
We substitute the deterministic velocity equation (\ref{1dotR}) into the stochastic 
Newtonian equation (\ref{2dyn-single-xi}), and derive the overdamped fluctuating components, 
\begin{eqnarray}\label{1dotR-xi}
\dot{\xi}^R_{x}&=&[\sigma_{xx}-\sigma^0_{xx}]{F^0_{x}}+[\sigma_{xy}-\sigma^0_{xy}]{F^0_{y}}\nonumber\\
&+&\sigma_{xx}{\xi}^F_{x}(t)+\sigma_{xy}{\xi}^F_{y}(t),\nonumber\\
\dot{\xi}^R_{y}&=&[\sigma_{yy}-\sigma^0_{yy}]{F^0_{y}}+[\sigma_{yx}-\sigma^0_{yx}]{F^0_{x}}\nonumber\\
&+&\sigma_{yx}{\xi}^F_{x}(t)+\sigma_{yy}{\xi}^F_{y}(t),
\end{eqnarray}
where the stochastic drifting tensor are
\begin{eqnarray}
\sigma_{xx}&=&\frac{{m}^{-1}\eta}{\eta^2+[{\omega}^0+{\xi}^{\omega}(t)]^2},\;\;\;
\sigma_{xy}=-\frac{{m}^{-1}[{\omega}^0+{\xi}^{\omega}(t)]}{\eta^2+[{\omega}^0+{\xi}^{\omega}(t)]^2},\nonumber\\
\sigma_{yy}&=&\frac{{m}^{-1}\eta}{\eta^2+[{\omega}^0+{\xi}^{\omega}(t)]^2},\;\;\;
\sigma_{yx}=\frac{{m}^{-1}[{\omega}^0+{\xi}^{\omega}(t)]}{\eta^2+[{\omega}^0+{\xi}^{\omega}(t)]^2},\nonumber\\
\end{eqnarray}
Combining the determinist position of the overdamped circle center and its fluctuating 
components around the deterministic trajectory, we get the instantaneous position of the particle itself in
the whole space,
\begin{eqnarray}
\tilde{x}(t)&=&\int\left\{\sigma_{xx}[F^0_{x}(t)+{\xi}^F_{x}(t)]+\sigma_{xy}[F^0_{y}(t)+{\xi}^F_{y}(t)]\right\}{dt}\nonumber\\
&+&[{r}_{0}(t)+{\xi}^r(t)]\cos[{\omega}^0+{\xi}^{\omega}(t)],\nonumber\\
\tilde{y}(t)&=&\int\left\{\sigma_{yy}[F^0_{y}(t)+{\xi}^F_{y}(t)]+\sigma_{yx}[F^0_{x}(t)+{\xi}^F_{x}(t)]\right\}{dt}\nonumber\\
&+&[{r}_{0}(t)+{\xi}^r(t)]\sin[{\omega}^0+{\xi}^{\omega}(t)]. 
\end{eqnarray}
Here the radii is fluctuating around an average value. For the special case, one may take the 
radii as constant ${r}_{0}(t)={r}_{0}=const$, the terms containing $\dot{r}_{0}(t)=0$ will become zero. 
The key difference between circling particle and non-circling particle is 
clearly shown in this equation. For a circling particle, 
a force in $X$-direction can induce the motion in $Y$-direction, 
and vice verse. For non-circling particle, the angular velocity is zero, 
$\omega=0$, i.e., $\sigma_{xy}=\sigma_{yx}=0$, the motion in $X$-direction and $Y$-direction are independent. 
It must be pointed out here that we do not put any constrain on the ratio between the determinist force term and 
the stochastic force term. If the stochastic force is much larger than deterministic force, the circle center 
would behave like Brownian particle.

The drifting of the circle center governs the probability distribution of the particle in large scale. 
If the center of the circle is fixed at one point, the particle is just circling around this center, 
the probability distribution along this circle is already known. We want to find the 
unknown probability distribution in the whole space. So we choose the size of the smallest unit box as much larger 
than the circle. Each value of the probability density function, $\psi$, is defined on one unit box. 
We define a probability flow: $\psi\dot{\vec{R}}$. The probability flow measures how fast the particle flows out of a small confined region. The decrease of 
the probability within this small confined region must equals to the integral of the probability flow along the boundary. 
This gives us the conservation equation of probability distribution, 
\begin{equation}\label{1conser}
\partial_t{\psi}+\nabla_R\cdot{(\psi\dot{\vec{R}})}=0.
\end{equation}

For a circling particle, we substitute Eq. (\ref{1dotR}) into Eq. (\ref{1conser}), the probability conservation equation reads
\begin{equation}\label{1difu2}
\frac{\partial\psi}{\partial{t}}+\sum_{ij}\partial_{Ri}[{\sigma}_{ij}{F}_{j}(x,y)\psi]=0.
\end{equation}
If the friction coefficient $\eta$ and drift tensor $\vec{\sigma}$ are state independent, we get a simpler diffusion equation,
\begin{eqnarray}\label{1difu3}
\frac{\partial\psi}{\partial{t}}
&+&[\sigma_{yy}F_{y}+\sigma_{yx}F_{x}]\;{\partial_{Ry}{\psi}}\nonumber\\
&+&[\sigma_{xx}F_{x}+\sigma_{xy}F_{y}]\;{\partial_{Rx}{\psi}}\nonumber\\
&+&[\sigma_{yy}\partial_{Ry}F_{y}+\sigma_{yx}\partial_{Ry}F_{x}]\;{{\psi}}\nonumber\\
&+&[\sigma_{xx}\partial_{Rx}F_{x}+\sigma_{xy}\partial_{Rx}F_{y}]\;{{\psi}}\nonumber\\
&=&0.
\end{eqnarray}
The drifting tensor for a circling particle are not arbitrary function, they must satisfy the relations,
\begin{eqnarray}
\sigma_{xx}&=&\sigma_{yy}=\frac{\eta}{\omega}\sigma_{yx},\;\;\;\;\;\sigma_{xy}=-\sigma_{yx}.
\end{eqnarray}
These constrains only exist for circling particles. The key physics of circling particle is lost in 
the conventional Fokker-Plank equation in which the diffusion tensor and drifting tensor
 are arbitrary independent functions.

We derived the probability conservation equation in polar coordinates following the same strategy 
above(See Appendix C for details), 
\begin{eqnarray}\label{1polar}
\frac{\partial{\psi}}{\partial{t}}
&+&\frac{\partial}{\partial{R}}[\sigma_{{\theta}{\theta}}{{\psi}}F_{{\theta}}+\sigma_{{\theta}{R}}{{\psi}}F_{{R}}
] \nonumber\\
&+&\frac{1}{R}\frac{\partial}{\partial{\theta}}
[\sigma_{{R}{R}}{{\psi}}F_{{R}}+\sigma_{{R}{\theta}}{{\psi}}F_{{\theta}}].\nonumber\\
&=&0.
\end{eqnarray}
In three dimensions, the probability conservation equation reads(detail calculation is presented in Appendix D),
\begin{eqnarray}
\frac{\partial{\psi}}{\partial{t}}
&+&[{\sigma_{xx}}F_{x}+{\sigma_{xy}}F_{y}+{\sigma_{xz}}F_{z}]\;\partial_{R_x}{{{\psi}}}\nonumber\\
&+&[{\sigma_{yx}}F_{x}+{\sigma_{yy}}F_{y}+{\sigma_{yz}}F_{z}]\;\partial_{R_y}{{{\psi}}}\nonumber\\
&+&[{\sigma_{zx}}F_{x}+{\sigma_{zy}}F_{y}+{\sigma_{zz}}F_{z}]\;\partial_{R_z}{{{\psi}}}\nonumber\\
&+&[{\sigma_{xx}}\partial_{R_x}F_{x}+{\sigma_{xy}}\partial_{R_x}F_{y}+{\sigma_{xz}}\partial_{R_x}F_{z}]\;{{{\psi}}}\nonumber\\
&+&[{\sigma_{yx}}\partial_{R_y}F_{x}+{\sigma_{yy}}\partial_{R_y}F_{y}+{\sigma_{yz}}\partial_{R_y}F_{z}]\;{{{\psi}}}\nonumber\\
&+&[{\sigma_{zx}}\partial_{R_z}F_{x}+{\sigma_{zy}}\partial_{R_z}F_{y}+{\sigma_{zz}}\partial_{R_z}F_{z}]\;{{{\psi}}}\nonumber\\
&=&0.
\end{eqnarray}
The tensor elements of the drifting tensor and diffusion tensor are not independent. As long as the 
angular velocity is not zero, the external force in 
$X$-direction $F_{x}$ will induce the motion in $Y-$ and $Z-$direction, 
so does the other force component, $F_{y}$ and $F_{z}$. This is the special dynamics of circling particles.

\section{The conservation equation of total number of many circling particles with overdamped circle center}

\begin{figure}[htbp]
\centering
\par
\begin{center}
\includegraphics[width=0.48\textwidth]{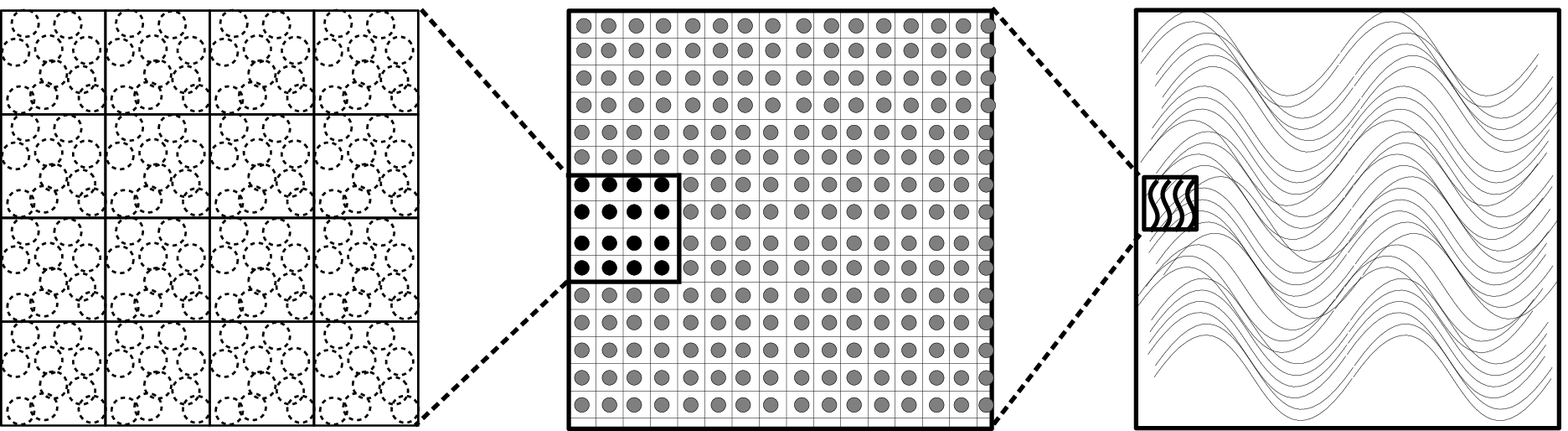}
\end{center}
\caption{\label{hydro} The viscous fluid model of many circling particles for the high density case. We first 
make coarse grain approximation of a number of circling particles confined by unit square, 
then take every unit square as one renormalized point particle. Finally, we condensed these point particles into compressible fluid.} \vspace{-0.2cm}
\end{figure}

We study the spatial distribution of many particle system and how this spatial distribution 
evolves as a function of time. We introduce a density distribution function ${N_{s}}$ which is only a 
function of space time, ${N_{s}}={N_{s}}(x,y,z,t)$. Density distribution quantifies the number 
of particles in an unit spatial area at certain time. We may choose two different spatial scales to define density.
One scale is the unit scale measuring the size of the particle itself. The other scale is the unit scale measuring the 
size of the circle. If one defines density as the number of particles in an unit scale of the particle itself, 
the density distribution is just a sharp peak moving along circular 
trajectory. The average of these peaks in a long time scale sit right at the center of the circle. When the center of 
the circle is fixed, the spatial distribution of particles large scale is almost static in spite of 
local fluctuation within every circle. The circle center dominates the spatial density distribution in large scale. 
Therefore the smallest unit scale of defining density should be larger than the radius of the circle(Fig. \ref{hydro}), 
\begin{equation}
{\Delta}x\geq{2{r}_{0}},\;\;{\Delta}y\geq{2{r}_{0}}, \;\;{\Delta}t\geq{{T}_{0}}. 
\end{equation}
where ${T}_{0}$ is the periodicity of circling motion.
The step size of experiment measurement must also satisfy the above constrain.

The density measures the number of particles in unit box ${\Delta}x{\Delta}y{\Delta}t$. 
Since the unit box ${\Delta}x{\Delta}y{\Delta}t$ is the smallest box we study, the relative motion of 
the particles inside the unit box does not contribute to the motion of the box. 
We take mean field approximation within every unit box. 
The collisions between particles within unit box is neglected. 
The center of mass of these ${N_{s}}$ particles sits 
right at this ideal point. The sum of the velocity of shifted to the lower half plane. The vortex sits in the highest 
density region.
these ${N_{s}}$ particles is the velocity of this ideal point. The centripetal force on the unit 
box is the sum of the centripetal force of the ${N_{s}}$ particles. The ideal point as a 
representation of unit box behaves like a renormalized particle with the total mass of ${N_{s}}$ 
particles. It interacts with other unit boxes like point particle interacting with point particle(Fig. \ref{hydro}). 
In this case, we take the many circling particle system as compressible fluid. The density function ${N_{s}}$ is governed 
by a pair of special hydrodynamic equations for compressible viscous fluid.

\subsection{The conservation equation of total particle number}

Diffusion is a non-equilibrium process to reach an equilibrium state. When many particles confined in a small 
region get the degree of freedom to move in a larger space, they diffuse into free space until the density distribution 
in the whole space becomes homogeneous. The entropy of the system increased during diffusion. 
The density gradient of inhomogeneous density distribution acts as an effective driving-force driving 
particles in high density region to low density region. This effective driving force is comprehensible 
from point view of the second law of thermodynamics. Entropy is increasing during diffusion.   
We assume the effective force of density gradient is 
$F_{gradient}=-\nabla_{R}N_{s}$. The particle is moving against high frictional resistance. 
The faster it moves forward, the stronger resistance it will encounter. In the most general cases, the resistant force non-linearly 
depends on velocity, 
${\arrowvert}F_{friction}{\arrowvert}=-\eta(\arrowvert\dot{\vec{R}}\arrowvert)^\beta$, $\beta$ is an index that could be real numbers. 
We only study the linear case for simplicity, $F_{friction}=-\eta\dot{\vec{R}}$.

The density distribution for the viscous fluid of circling particles is governed by fluid dynamic equation and 
particle number conservation equation, 
\begin{eqnarray}\label{dyn-single}
&&mN_{s}\ddot{\vec{R}}
=-mN_{s}\dot{\vec{R}}\times{\vec{\omega}}+N_{s}\vec{F}-D\nabla{N_{s}}-mN_{s}\eta\dot{\vec{R}},\nonumber\\
&&\partial_t{{N}_{s}}+\nabla_R\cdot({N_{s}\dot{\vec{R}}})=0.
\end{eqnarray}
Here $\eta$ is the friction coefficient. The diffusion force term from density gradient 
only exist in many particle system. The density distribution of a single particle is a delta function at one point, 
there is no density gradient. The force term in Eq. (\ref{dyn-single}) is the sum of a deterministic force and stochastic force,
\begin{eqnarray}
F_{x}&=&F^0_{x}(t)+{\xi}^F_{x}(t),\;\;\;\;F_{y}={F}^0_{y}(t)+{\xi}^F_{y}(t). 
\end{eqnarray}
The angular velocity is fluctuating around central value $\omega_0$,
\begin{eqnarray}
{\omega}&=&{\omega}^0+{\xi}^{\omega}(t). 
\end{eqnarray}
The coarse grain approximation take all the particles in the smallest unit box as one mean-field particle, it is no longer 
important to consider the relative positions between particles within the smallest unit box. To show the difference between single 
particle dynamics and many particle dynamics, we divide both sides of the 
fluid dynamic equations (\ref{dyn-single}) by density function to derive the mean-field velocity of the many particles 
within one unit box, 
\begin{eqnarray}\label{dynamic2}
m\ddot{R}_{x}&=&-m{\omega}\dot{R}_{y}-D_x\frac{\partial_{Rx}{N_{s}}}{N_{s}}-m\eta\dot{R}_{x}+{F}_{x},\nonumber\\
m\ddot{R}_{y}&=&m{\omega}\dot{R}_{x}-D_y\frac{\partial_{Ry}{N_{s}}}{N_{s}}-m\eta\dot{R}_{y}+{F}_{y}.
\end{eqnarray}
Noticing here the velocity is not the velocity of a single particle, but is the velocity of the coarse grained particle which represents 
all the particles within the unit box. The diffusion coefficient $D_x$ and $D_y$ are state dependent function, $D_x=D_x(x,y)$ and $D_y=D_y(x,y)$. 
The friction force exerted on all the circle centers is much larger than the inertial term. 
The motion of the coarse grain particle is overdamped. We derived the overdamped velocity of the unit box 
from Eq. (\ref{dynamic2}) by imposing $m\ddot{R}_{x}=0,\;\;\;m\ddot{R}_{y}=0,$
\begin{eqnarray}\label{vxy2}
\dot{R}_{x}&=&-{D_{xy}}\frac{\partial_{Ry}{N_{s}}}{N_{s}}-
{D_{xx}}\frac{\partial_{Rx}{N_{s}}}{N_{s}}+\sigma_{xx}F_{x}+\sigma_{xy}F_{y},\nonumber\\
\dot{R}_{y}&=&-{D_{yx}}\frac{\partial_{Rx}{N_{s}}}{N_{s}}
-{D_{yy}}\frac{\partial_{Ry}{N_{s}}}{N_{s}}+\sigma_{yy}F_{y}+\sigma_{yx}F_{x},\nonumber\\
\end{eqnarray}
where the diffusion tensor $D$ and drifting tensor $\sigma$ are
\begin{eqnarray}
D_{xx}&=&\frac{{D}_{x}\eta}{mA^{2}},\;\;\;\;\;\;D_{xy}=-\frac{D_{y}\omega}{mA^{2}},\nonumber\\
D_{yy}&=&\frac{{D}_{y}\eta}{mA^{2}},\;\;\;\;\;\;D_{yx}=\frac{D_{x}\omega}{mA^{2}},\nonumber\\
\sigma_{xx}&=&\frac{\eta}{mA^{2}},\;\;\;\;\;\;\;\sigma_{xy}=-\frac{\omega}{mA^{2}},\nonumber\\
\sigma_{yy}&=&\frac{\eta}{mA^{2}},\;\;\;\;\;\;\;\sigma_{yx}=\frac{\omega}{mA^{2}},
\end{eqnarray}
$A^{2}=\eta^2+\omega^2$ is a normalization factor. The flow of the renormalized particles is
\begin{eqnarray}
J_{i}=N_{s}\dot{R}_{i}=\sum_j\left[{\sigma}_{ij}{F}_{j}(x,y)N_{s}-{D}_{ij}(x,y){\partial_{j}{N}_{s}}\right].
\end{eqnarray}
The velocity $\dot{R}_{i}$ is the velocity of unit box with mean field approximation inside the box. 
We study how many particles is distributed in the certain area at time $t$. 
The unit box is much larger than the radius of the circle. If the centers of the particles inside the unit box are fixed, 
no matter how fast they are circling around their center, the density function always has the same value. 
So we neglect the local circling motion inside the unit box, and only take into account of the motion of the circle center. 
If we draw an arbitrary closed curve to confine a number of unit boxes, the number of particles flowing 
out of this close curve equals the integration of the decreased density function 
covering the area enclosed by this close curve. The conservation equation of total particle number expressed by flow current is 
\begin{equation}
\partial_t{{N}_{s}}+\nabla_R\cdot{\textbf{J}}=0,\;\;\;\;\;\;\; \textbf{J}={N_{s}\dot{\vec{R}}}
\end{equation} 
The first term of this conservation equation means the decrease of total number of particles in this area, 
the second term counts how many particles flow out of this area by crossing the border. 
Substituting velocity Eq. (\ref{vxy2}) into the conservation equation, we derived the evolution equation 
of density distribution,
\begin{equation}\label{difu2}
\frac{\partial{N}_{s}}{\partial{t}}+\sum_{ij}\partial_{Ri}[{\sigma}_{ij}{F}_{j}(x,y){N}_{s}-{D}_{ij}(x,y){\partial_{Rj}{N}_{s}}]=0.
\end{equation}
In the most general case, both the friction coefficient $\eta$  and the drift tensor are taken 
as state dependent function, $\vec{\sigma}=\vec{\sigma}(x,y)$, $\eta=\eta(x,y)$. 
If diffusion tensor and drift tensor are both constant, the explicit formulation of diffusion equation reads, 
\begin{eqnarray}\label{difu3}
\frac{\partial{N}_{s}}{\partial{t}}
&-&[D_{xy}+D_{yx}]\partial_{Rx}\partial_{Ry}N_{s}\nonumber\\
&-&D_{xx}\partial_{Rx}^{2}N_{s}-D_{yy}\partial_{Ry}^{2}N_{s} \nonumber\\
&+&[\sigma_{yy}F_{y}+\sigma_{yx}F_{x}]\;{\partial_{Ry}{N_{s}}}\nonumber\\
&+&[\sigma_{xx}F_{x}+\sigma_{xy}F_{y}]\;{\partial_{Rx}{N_{s}}}\nonumber\\
&+&[\sigma_{yy}\partial_{Ry}F_{y}+\sigma_{yx}\partial_{Ry}F_{x}]\;{{N_{s}}}\nonumber\\
&+&[\sigma_{xx}\partial_{Rx}F_{x}+\sigma_{xy}\partial_{Rx}F_{y}]\;{{N_{s}}}\nonumber\\
&=&0.
\end{eqnarray}
If there is no external field, $F_{x}=0,F_{y}=0$, the diffusion equation becomes
\begin{eqnarray}\label{difu4}
\frac{\partial{N}_{s}}{\partial{t}}
&-&[D_{xy}+D_{yx}]\partial_{Rx}\partial_{Ry}N_{s}\nonumber\\
&-&D_{xx}\partial_{Rx}^{2}N_{s}-D_{yy}\partial_{Ry}^{2}N_{s} =0.
\end{eqnarray}
This equation demonstrated the special character of many particle system.
 In the probability conservation equation of single circling particle, if the external field is zero, 
all the special dynamic of circling particle vanished, there is no difference between 
a circling particle and a non-circling particle in large scale. While for many circling particle system, 
even if there is no any external field, the special diffusing dynamics of circling particle still exist.

We also get the conservation equation of many circling particles in polar coordinates system 
(see Appendix E for detail calculation), 
\begin{eqnarray}\label{polar}
\frac{\partial{N}_{s}}{\partial{t}}
&-&\frac{\partial}{\partial{R}}
[{D_{{R}{\theta}}}\frac{1}{R}\frac{\partial}{\partial{\theta}}{{N_{s}}}+{D_{{R}{R}}}{\frac{\partial}{\partial{R}}{N_{s}}}]\nonumber\\
&+&\frac{\partial}{\partial{R}}[\sigma_{{\theta}{\theta}}{N_{s}}F_{{\theta}}+\sigma_{{\theta}{R}}{N_{s}}F_{{R}}
] \nonumber\\
&-&\frac{1}{R}\frac{\partial}{\partial{\theta}}
[{D_{{\theta}{R}}}{\frac{\partial}{\partial{R}}{N_{s}}}+
{D_{{\theta}{\theta}}}\frac{1}{R}\frac{\partial}{\partial{\theta}}{{N_{s}}}] \nonumber\\
&+&\frac{1}{R}\frac{\partial}{\partial{\theta}}
[\sigma_{{R}{R}}{N_{s}}F_{{R}}+\sigma_{{R}{\theta}}{N_{s}}F_{{\theta}}]\nonumber\\
&=&0.
\end{eqnarray}
The three dimensional evolution 
equation of density distribution is(see Appendix F for detail calculation)
\begin{equation}\label{difu3D}
\frac{\partial{N}_{s}}{\partial{t}}+\sum_{ij}\frac{\partial}{{\partial{R}}_i}[{\sigma}_{ij}{F}_{j}N_{s}]
-\sum_{ij}\frac{\partial}{{\partial{R}}_i}[{D}_{ij}\frac{\partial}{{\partial{R}}_j}{N}_{s}]=0.
\end{equation}
here the friction coefficient $\eta=\eta(R_x,R_y,R_z)$, drift tensor ${\sigma}_{ij}=\sigma(R_x,R_y,R_z)$ 
and diffusion tensor $D_{i}=D_{i}(R_x,R_y,R_z)$ are all state-dependent. 
The detail expression of drifting tensor ${\sigma}_{ij}$ and diffusion tensor ${D}_{ij}$ are presented in Appendix F.

\subsection{Diffusion tensor with fluctuating angular frequency}

\begin{figure}[htbp]
\centering 
\par
\begin{center}
$
\begin{array}{c@{\hspace{0.03in}}c}
%\multicolumn{1}{1}{\mbox{}} & \multicolumn{1}{1}{\mbox{}} \\
\includegraphics[width=0.24\textwidth]{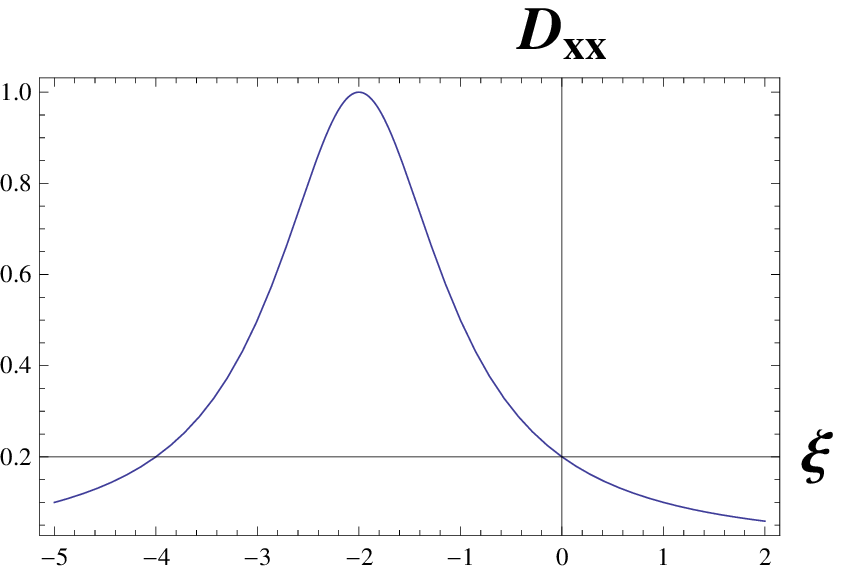}& \includegraphics[width=0.24\textwidth]{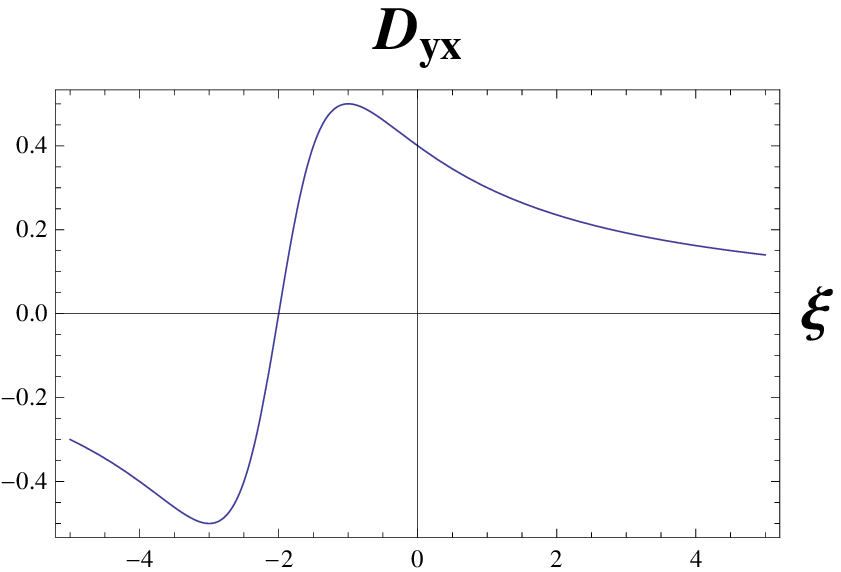}\\
\mbox{(a)} & \mbox{(b)}\\
\end{array}
$
\end{center}
\caption{\label{Dxx} (a) The diagonal diffusion coefficients ($D_{xx}$=$D_{xy}$) under 
random fluctuation $\xi$. We have set   $\eta/m=1$, $\omega=2$.
 (b) The off-diagonal diffusion coefficients ($D_{xy}$=$-D_{yx}$) as a 
function of random fluctuation $\xi$. Parameters are given by $\eta/m=1$, $\omega=2$.} \vspace{-0.2cm}
\end{figure}

There are three independent sources that may lead to a fluctuating circular trajectory: the fluctuating center, 
the fluctuating radius and the fluctuating angular frequency(see Appendix B for details). We study how 
the fluctuation of angular frequency modify the diffusion tensor in this section. 
The diffusion tensor is an algebra equation of friction coefficient, diffusion coefficient and angular velocity. 
If the circling speed along the circle is not constant, 
we introduce a fluctuating variable $\xi^{\omega}(t)$ into the angular velocity, $\omega=\omega^0+\xi^{\omega}(t),$ 
where $\omega^0$ is the central angular frequency. We take the average of the fluctuation 
within one periodicity, $\langle\xi^{\omega}(t)\rangle={\xi}$. 
The diffusion tensor with an average amplitude of fluctuations reads 
\begin{eqnarray}\label{Dxxx}
D_{xx}&=&\frac{D_x\eta}{m([\omega^0+{\xi}]^2+\eta^2)},\nonumber\\
D_{yx}&=&\frac{D_x[\omega^0+{\xi}]}{m([\omega^0+{\xi}]^2+\eta^2)}.
\end{eqnarray}
If the fluctuation reduced the angular velocity, both diffusion coefficient $D_{xx}$ and $D_{xy}$ will 
increase(Fig. \ref{Dxx}), but the  $D_{yx}$ will reach climax first(Fig. \ref{Dxx} (b)).
 If the fluctuation increase the angular velocity, both the diffusion coefficients 
$D_{xx}$ and $D_{xy}$ will decrease, $D_{xy}$ decreases much slower than $D_{xx}$. 
The ratio between $D_{xx}$ and $D_{xy}$ is,
\begin{equation}
\frac{D_{xx}}{D_{yx}}=\frac{\eta}{\omega^0+\xi}.
\end{equation}
The increase of frequency will make the off-diagonal diffusion stronger. 
As the total number of particle is conserved, the increasing of off-diagonal diffusion will 
result in the decreasing of the diagonal diffusion.

\subsection{Effective force field induced by velocity gradient}

In the microscopic picture, the fast moving particle has large momentum to push the slow moving particles. 
There exist effective force on the boundary between fast moving particles and slow moving particles.
One special character of a circling particle is its local circling velocity in different directions 
are not independent, i.e., $(\dot{r}_{x})^2+(\dot{r}_{y})^2=(r_0)^2\omega^2.$ 
The velocity gradient between different directions are strongly correlated. 
The velocity gradient in one direction can induce a effective force in its perpendicular direction. 
We define an effective force field induced by velocity gradient,
\begin{equation}\label{J} 
\vec{J}=-\nabla{V}{\times}\vec{\omega},
\end{equation}
where $\vec{\omega}=\omega\textbf{e}_z$ is the normal vector perpendicular to $X-Y$ plane. 
$\omega<0$ means the swimmer is rotating in clockwise direction. 
$\omega>0$ means the circle swimmer rotates in counterclockwise direction. 
The derivative of velocity with respect to time has the dimensionality of acceleration. 
According to Newton's law, the vector field $\vec{J}$ is an effective force field. 
For the velocity field of circling particle, the force field is
\begin{eqnarray}\label{Jxy}
J_{y}&=&\omega\partial_{Rx}[\sigma_{xx}F_{x}]+\omega\partial_{Rx}[\sigma_{xy}F_{y}],\nonumber\\
&+&{\omega}\partial_{Rx}[D_{xy}{\partial_{Ry}{\ln}{N_{s}}}]+
\omega}\partial_{Rx}[{D_{xx}{\partial_{Rx}{\ln}{N_{s}}}]\nonumber\\
J_{x}&=&\omega\partial_{Ry}[\sigma_{yy}F_{y}]
+\omega\partial_{Ry}[\sigma_{yx}F_{x}].\nonumber\\
&+&{\omega}\partial_{Ry}[D_{yx}{\partial_{Rx}{\ln}{N_{s}}}]+
{\omega}\partial_{Ry}[D_{yy}{\partial_{Ry}{{\ln}N_{s}}}].
\end{eqnarray}
If the circling particle switch its circling direction, i.e., 
from clockwise to counterclockwise, or vice verse,
any local vector in the force field switch its direction too. 
The switching of circling direction is denoted by flipping the sign 
of the angular velocity, $\omega\rightarrow-\omega$, or  $-\omega\rightarrow\omega$. 
This point is clearly shown in Eq. (\ref{Jxy}).

\begin{figure}[htbp]
\centering 
\par
\begin{center}
$
\begin{array}{c@{\hspace{0.06in}}c}
%\multicolumn{1}{1}{\mbox{}} & \multicolumn{1}{1}{\mbox{}} \\
\includegraphics[width=0.16\textwidth]{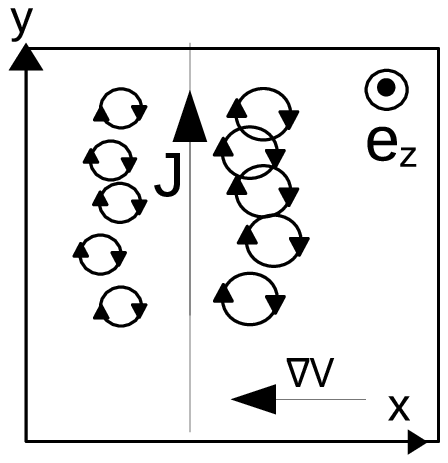}& \includegraphics[width=0.16\textwidth]{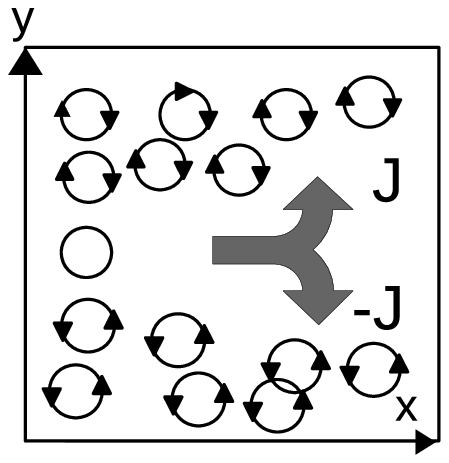}\\
\mbox{(a)} & \mbox{(b)}\\
\end{array}
$
\end{center}
\caption{\label{flow} (a) The transverse flow induced by velocity gradient. (b) Swimmers with opposite circling handedness will split during diffusion.} \vspace{-0.2cm}
\end{figure}

The velocity gradient in X-direction will generate effective force in Y-direction(Fig. \ref{flow} (a)). 
If there are two groups of particles, and the circling direction in one group 
is opposite to the other group. If one mix the two groups together, and let them 
diffuse in a rectangular channel. The particles will split into two 
separated groups during diffusion. One group goes to positive Y-direction, 
the other goes to negative Y-direction (Fig. \ref{flow} (b)).

\subsection{Vortex in diffusion phenomena}

Vortex is nonlinear collective phenomena of many body system. For any given vector field, one can use 
vorticity to quantify vortex. The vorticity of the circle center's velocity field Eq. (\ref{vxy2}) is    
\begin{equation}\label{vortex2}
\Gamma_V=\vec{\nabla}{\times}\vec{V}=\partial_{Rx}\dot{R}_{y}-\partial_{Ry}\dot{R}_{x}. 
\end{equation}
We first consider a special case that the diffusion tensor are state-independent, 
and the diffusion coefficient in X-direction equals to the diffusion coefficient in Y-direction, 
${D}_{x}={D}_{y}$. The diffusion tensor and drifting tensor satisfy the relations, 
\begin{eqnarray}
D_{xx}&=&D_{yy},\;\;D_{xy}=-D_{yx},\nonumber\\
\sigma_{xx}&=&\sigma_{yy},\;\;\;\;\;\sigma_{xy}=-\sigma_{yx}.
\end{eqnarray}
Substituting Eq. (\ref{vxy2}) into Eq. (\ref{vortex2}) give us
\begin{eqnarray}\label{LV}
\Gamma_V&=&\sigma_{xx}(\partial_{Rx}F_{y}-\partial_{Ry}F_{x})
+\sigma_{yx}(\partial_{Rx}F_{x}+\partial_{Ry}F_{y})\nonumber\\
&+&{D_{xy}}(\partial_{Ry}{\partial_{Ry}}{\ln}{N_{s}}+\partial_{Rx}{\partial_{Rx}}{\ln}{N_{s}}).
\end{eqnarray}
According to Green-function theory, 
\begin{equation}
\partial_{Ry}{\partial_{Ry}}{\ln}{N_{s}}+\partial_{Rx}{\partial_{Rx}}{\ln}{N_{s}}=2\pi\delta(N_{s}),
\end{equation}
we simplify $\Gamma_V$, 
\begin{eqnarray}\label{gammaV}
\Gamma_V&=&\sigma_{xx}(\partial_{Rx}F_{y}-\partial_{Ry}F_{x})
+\sigma_{yx}(\partial_{Rx}F_{x}+\partial_{Ry}F_{y})\nonumber\\
&+&{D_{xy}}2\pi\delta(N_{s}).
\end{eqnarray}
When the particles are circling around a common center, 
they are pulled away from the center by centrifugal force, 
so the density vanishes at the center. 
More over, the force field induced by velocity gradient may also 
have vortex configuration. We measure this kind of vortex by
\begin{equation}
\Gamma_J=\vec{\nabla}{\times}\vec{J}=\partial_{Rx}J_{y}-\partial_{Ry}J_{x},
\end{equation}
Both $\Gamma_J$ and $\Gamma_V$ are determined by the density 
distribution and external field. One can technically split the 
density function into two components, and express the density 
function as ${N_{s}}=\sqrt{N^2_{a}+N^2_{b}}.$ Then every vortex can 
be identified by a topological number according to Duan's topological 
current theory\cite{duan}. The vortex configuration evolute 
following the evolution of density distribution.

\section{Numerical evolution of the density distribution for many circling particles in two dimensions}

The evolution equation of density distribution (\ref{difu2}) in two dimensions is a nonlinear equation. 
Exact solution only exist in some special cases. We performed some numerical computation to study 
the density distribution, vortex field and force field. The numerical results are in 
good agreement with our theoretical prediction.

\subsection{The evolution of density distribution for $D_x={D}_y$}

We first study a special case of diffusion equation (\ref{difu2}). 
The diffusion tensor and drifting tensor are state-independent. 
The diffusion coefficient in the two directions are chosen as equal, $D_x=D_y=D.$ 
The off-diagonal diffusion coefficients and drifting coefficients are, 
\begin{equation}\label{of}
\sigma_{yx}=-\sigma_{xy}=\frac{\omega}{mA^{2}},\;\;D_{yx}=-D_{xy}=\frac{D\omega}{mA^{2}}.
\end{equation}
A constant force in X-direction is applied and no external force exists in Y-direction, 
\begin{equation}
F_{x}=C_{x}, \;\;\;\;F_{y}=0.
\end{equation}
For a non-circling particle, if $F_{y}=0$, the density center will not drift in Y-direction. 
But for circling particles, even if $F_{y}=0$, the density center still has a drifting velocity in Y-direction. 
The diffusion equation (\ref{difu2}) for this special case reads
\begin{equation}\label{sperm}
\frac{\partial{N}_{s}}{\partial{t}}=D_{xx}\partial_{Rx}^{2}N_{s}+D_{yy}\partial_{Ry}^{2}N_{s}
-b\;{\partial_{Ry}{N_{s}}}-a\;{\partial_{Rx}{N_{s}}}.
\end{equation}
We take the boundary as infinite where the density decays to zero. 
The analytic solution of density distribution under this boundary condition is 
\begin{eqnarray}\label{solut1}
N_{s}=\frac{1}{4\pi{D_{r}}t}\exp[-\frac{(R_x-at)^2}{4D_{r}t}-\frac{(R_y-bt)^2}{4D_{r}t}].
\end{eqnarray}
This solution describes a propagating Gaussian packet. 
The drifting speed of the density center in $X$-direction is $a=\sigma_{xx}F_{x}$. 
The drifting velocity of density center along Y-direction is $b=\sigma_{yx}F_{x}$. 
In mind of the fact that there is no external driven force in 
Y-direction, this drifting velocity along Y-direction is a special character 
of circling particles.

\begin{figure}[htbp]
\centering
\par
\begin{center}
\includegraphics[width=0.32\textwidth]{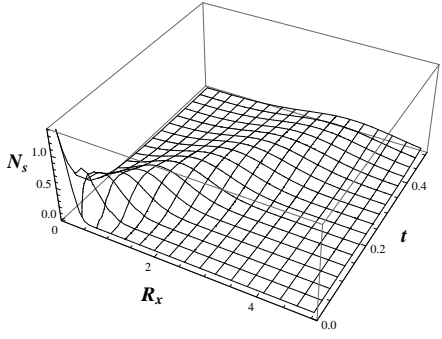}
\end{center}
\caption{\label{exama}  The density evolution of solution equation (\ref{solut1}) for $a/D_r=7$. The drifting velocity is larger than diffusion velocity, so one can observe a moving pump.  } \vspace{-0.2cm}
\end{figure}

\begin{figure}[htbp]
\centering
\par
\begin{center}
\includegraphics[width=0.32\textwidth]{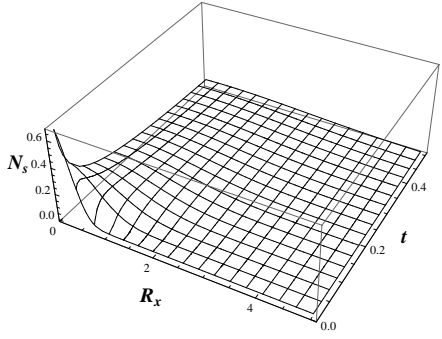}
\end{center}
\caption{\label{examb} The density evolution of solution equation (\ref{solut1}) 
for $a/D_r=1/4$. The drifting speed is slower than the diffusion speed. } \vspace{-0.2cm}
\end{figure}

We integrated the density along Y-axis to derive the density evolution 
along X-axis,  $N_{s}(R_x,t)=\int_0^{h}N_{s}dR_y$, where $h$ is the width of channel 
along Y-direction. If the drifting velocity is much larger than the diffusion coefficient, 
there appears apparent drifting of the density center(Fig. \ref{exama}). If the diffusion 
coefficient is much larger than the drifting, the density distribution shows 
a rapid decay in beginning, then there appears a plateau due to the drifting of density center 
(Fig. \ref{examb}). We also integrated $R_x$ out to keep the $N_{s}$ as a function 
of $(R_y,t)$, $N_{s}(R_y,t)=\int_0^{h'}N_{s}dR_x$. 
The evolution of density distribution shows very similar behavior as above. In fact, the solution equation (\ref{solut1})
 is symmetric is we exchange the two variables $x$ and $y$, and exchange $a$ and $b$.

\begin{figure}[htbp]
\centering
\par
\begin{center}
\includegraphics[width=0.27\textwidth]{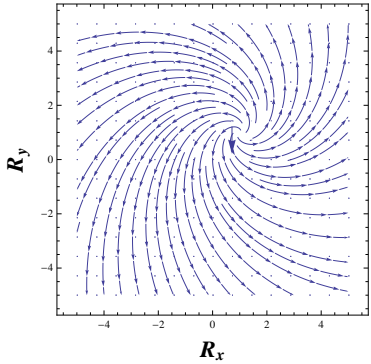}
\end{center}
\caption{\label{t13} The velocity field $\vec{V}=(V_{x},V_{y})$ for density solution (\ref{solut1}) at time $t=1$. The drifting parameters are a=1, b=1. Diffusion coefficient $D_r=1$.} \vspace{-0.2cm}
\end{figure}

The velocity field $\vec{V}_R=(V_{x},V_{y})$ for the analytic density 
distribution Eq. (\ref{solut1}) demonstrates a typical vortex configuration(Fig. \ref{t13}). 
The vorticity of the density distribution function Eq. (\ref{solut1}) is obtained by substituting velocity 
Eq. (\ref{vxy2}) into the vorticity Eq. (\ref{vortex2})
\begin{eqnarray}\label{vorR}
\Gamma_V={D_{xy}}[\partial_{Ry}{\partial_{Ry}}L({R})+\partial_{Rx}{\partial_{Rx}}L({R})],
\end{eqnarray}
where the function $L({R})$ is
\begin{eqnarray}\label{LR}
L({R})=\left[-\frac{(R_x-at)^2}{4D_{r}t}-\frac{(R_y-bt)^2}{4D_{r}t}\right]^{-4\pi{D_{r}}t}.
\end{eqnarray}
The plot of high vorticity region with respect to the vortex 
configuration of Fig. \ref{t13} is a small disc region covered the center of the vortex.

\subsection{The evolution of density distribution for $D_x\neq{D}_y$}

When $D_x\neq{D}_y$, the diffusion equation will include a cross-diffusion term, 
$D_{yx}\partial_{Rx}\partial_{Ry}N_{s}$.This cross-diffusion term would vanish for zero angular velocity. 
We first study the special case of a positive angular frequency, $\omega>0$, in this section. 
The diffusion coefficients and drifting tensor are state independent. 
The external force is constant,
\begin{equation}
F_{x}=C_{x},\;\;F_{y}=C_{y}.
\end{equation} 
The diffusion equation for $D_x{\neq}D_y$ is,
\begin{eqnarray}\label{sperm2}
\frac{\partial{N}_{s}}{\partial{t}}&=&
D_{xx}\partial_{Rx}^{2}N_{s}+D_{yy}\partial_{Ry}^{2}N_{s}\nonumber\\
&+&(D_{yx}+D_{xy})\partial_{Rx}\partial_{Ry}N_{s}\nonumber\\
&-&b\;{\partial_{Ry}{N_{s}}}-a\;{\partial_{Rx}{N_{s}}}.
\end{eqnarray}
The drifting velocity of density center is $a=\sigma_{xx}C_{x}+\sigma_{xy}C_{y}$, $b=\sigma_{yy}C_{y}+\sigma_{yx}C_{x}$.

\begin{figure}[htbp]
\centering
\par
\begin{center}
$
\begin{array}{c@{\hspace{0.03in}}c}
%\multicolumn{1}{1}{\mbox{}} & \multicolumn{1}{1}{\mbox{}} \\
\vspace{-0.1cm}\hspace{-0.4cm}
\includegraphics[width=0.25\textwidth]{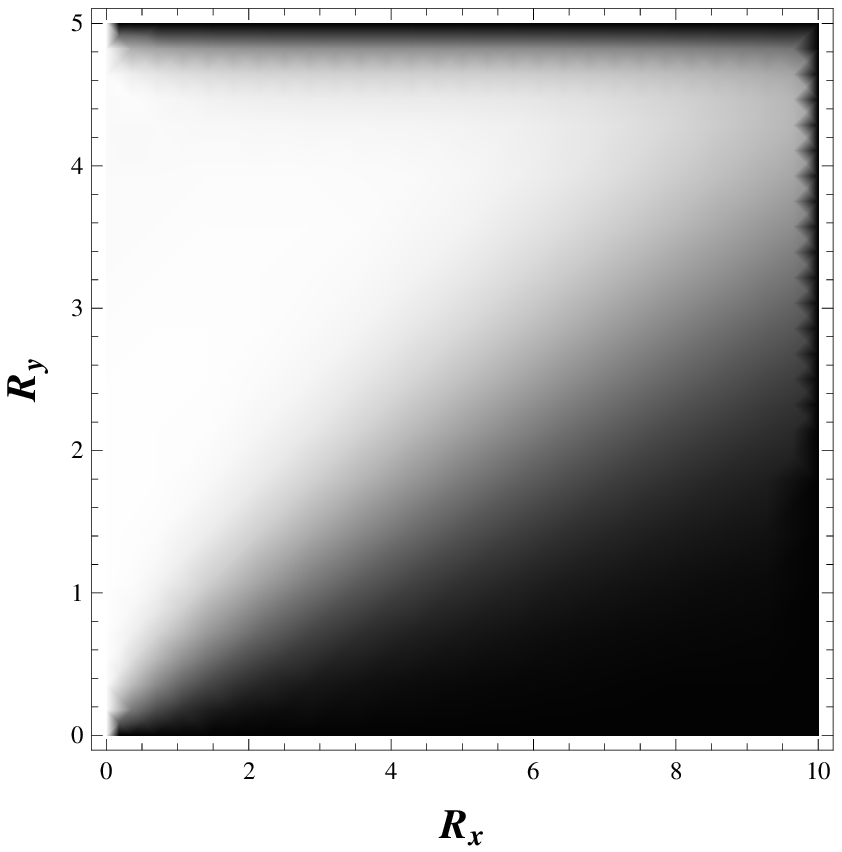}&\includegraphics[width=0.25\textwidth]{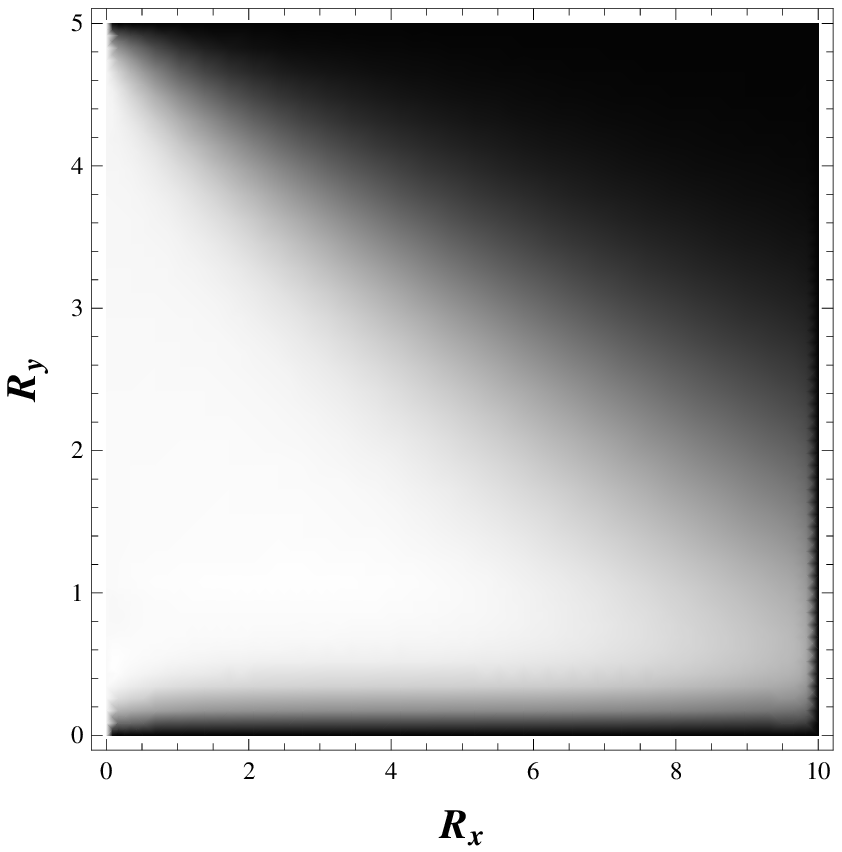}\\
\mbox{(a)}& \mbox{(b)}\\
\vspace{-0.1cm}\hspace{-0.4cm}
\includegraphics[width=0.25\textwidth]{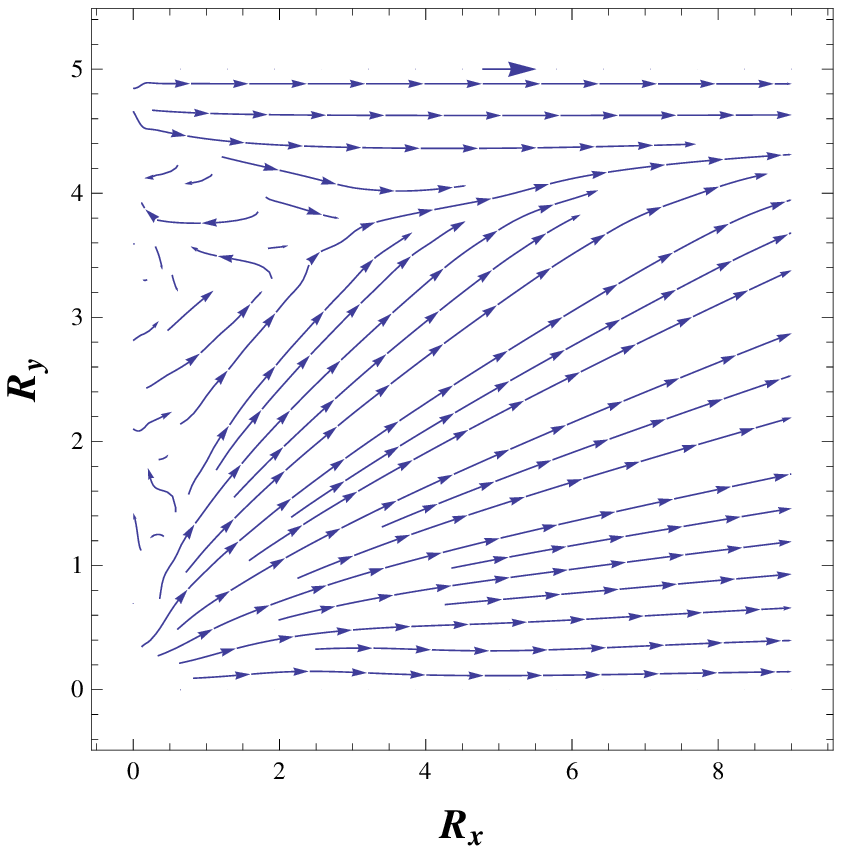}&\includegraphics[width=0.25\textwidth]{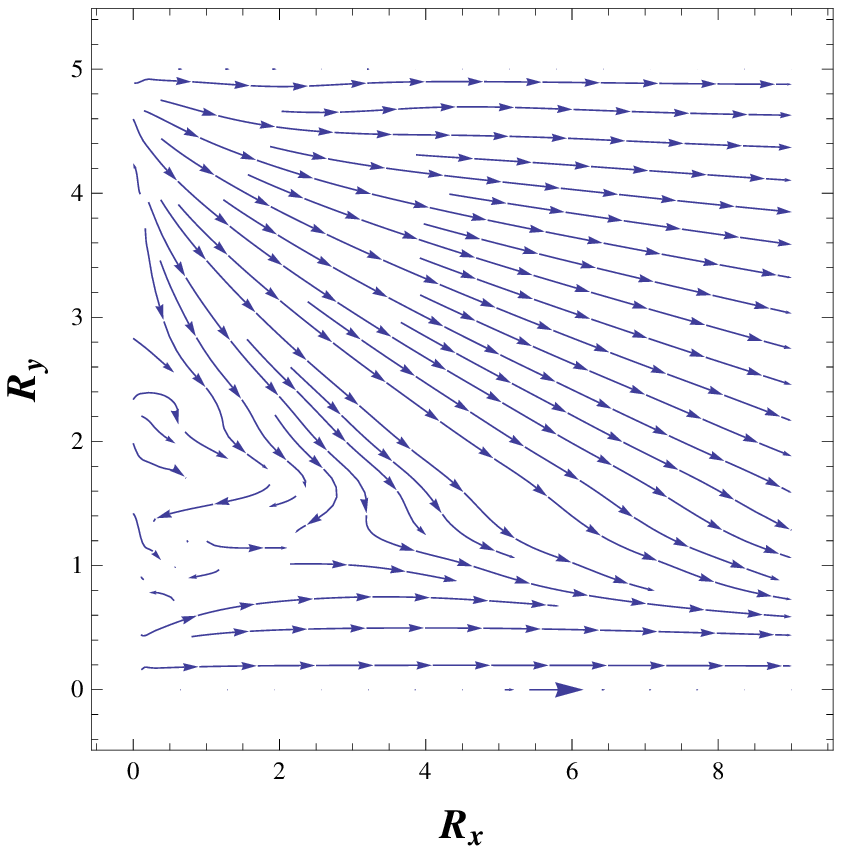} \\
\mbox{(c)}& \mbox{(d)}\\
\end{array}
$
\end{center}
\caption{\label{difu} (a)  The density distribution ${N}_{s}({Rx},{Ry})$ at time $t=15$ for 
the numerical evolution of diffusion equation (\ref{sperm2}). The bright color indicates high density, 
dark color indicates low density. The particles concentrate on the line $R_x=0$ at time $t=0$. 
parameters are $D_{yx}+D_{xy}=-1$, $\omega>0$, $b=5$ and $a=15$. (b) One snapshot of the numerical 
evolution of density distribution of equation (\ref{sperm2}) at time $t=15$, $\omega<0$. 
The bright color indicates high density, dark color indicates low density. (c) 
The effective force field $\vec{J}=(J_{x},J_{y})$ corresponding to the density distribution of 
Fig. \ref{difu} (a), $\omega>0$. (d) The transverse force field $\vec{J}=(J_{x},J_{y})$ for $\omega<0$ with respect to
 the density distribution of Fig. \ref{difu} (b). All the parameters are the same
 as that for Fig. \ref{difu} (b).} 
\vspace{-0.2cm}
\end{figure}

A numerical computation of density distribution is performed in two dimensions. In the beginning, the particles 
are confined along a line parallel to Y-axes at ${R_x}=0$. 
The initial density is ${N}_{s}=10$. The diffusion coefficients are $D_{xx}=0.9$ and $D_{yy}=1$. We choose a large 
drifting velocity, $b=5$ and $a=15$. The off-diagonal diffusion coefficient is 
$D_{yx}+D_{xy}=-1$ which corresponds to a positive angular frequency. One snapshot of 
the density evolution at $t=15$ is shown in Fig. \ref{difu} (a). The light color demonstrates
 the high density. The dark color shows the low density. The center of high 
density region shifts to positive ${Ry}$-direction(Fig. \ref{difu} (a)). 
Later we increased the off-diagonal diffusion tensor and drift tensor 
to $D_{yx}+D_{xy}=-2$, b=10, while keep the diagonal diffusion tensor as the same as before. 
A snapshot of density evolution at $t=15$ shows that the drift in Y-direction speeds 
up. While the drifting of density center in X-direction slows down.  
.

\begin{figure}[htbp]
\centering
\par
\begin{center}
$
\begin{array}{c@{\hspace{0.03in}}c}
%\multicolumn{1}{1}{\mbox{}} & \multicolumn{1}{1}{\mbox{}} \\
\vspace{-0.1cm}\hspace{-0.4cm}
\includegraphics[width=0.25\textwidth]{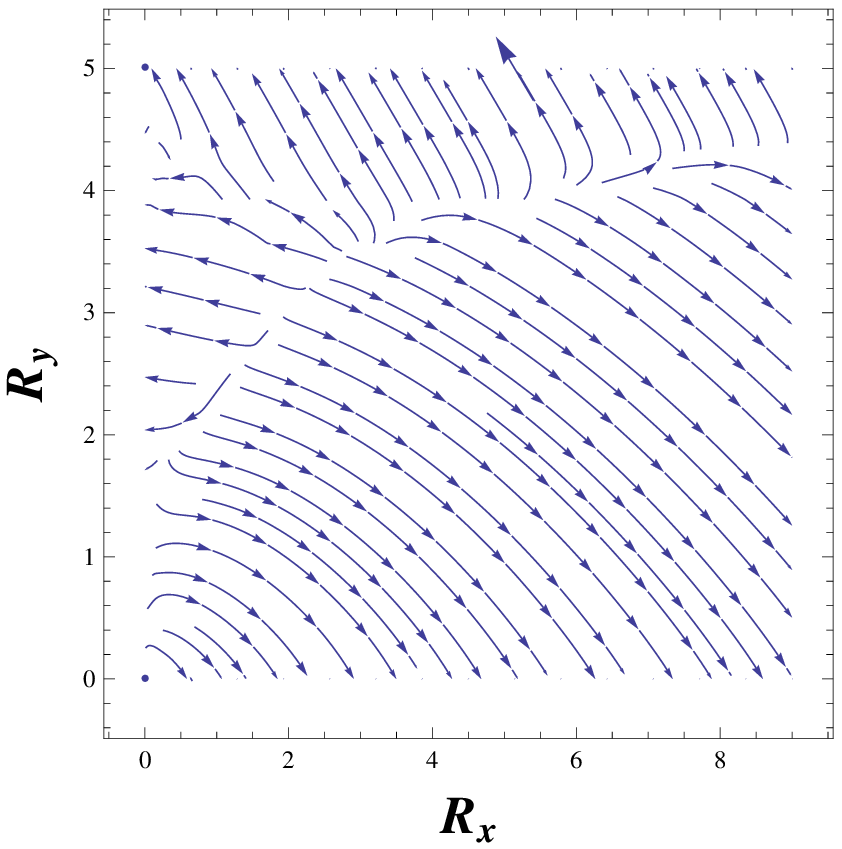}&\includegraphics[width=0.25\textwidth]{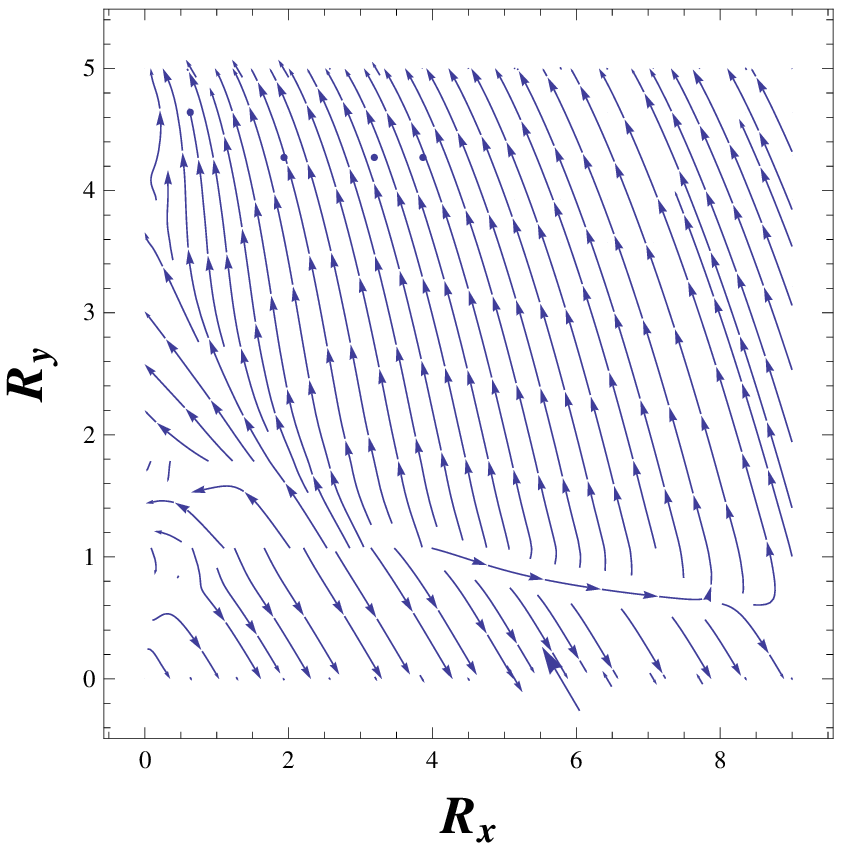} \\
\mbox{(a)}& \mbox{(b)}\\
\vspace{-0.1cm}\hspace{-0.4cm}
\includegraphics[width=0.25\textwidth]{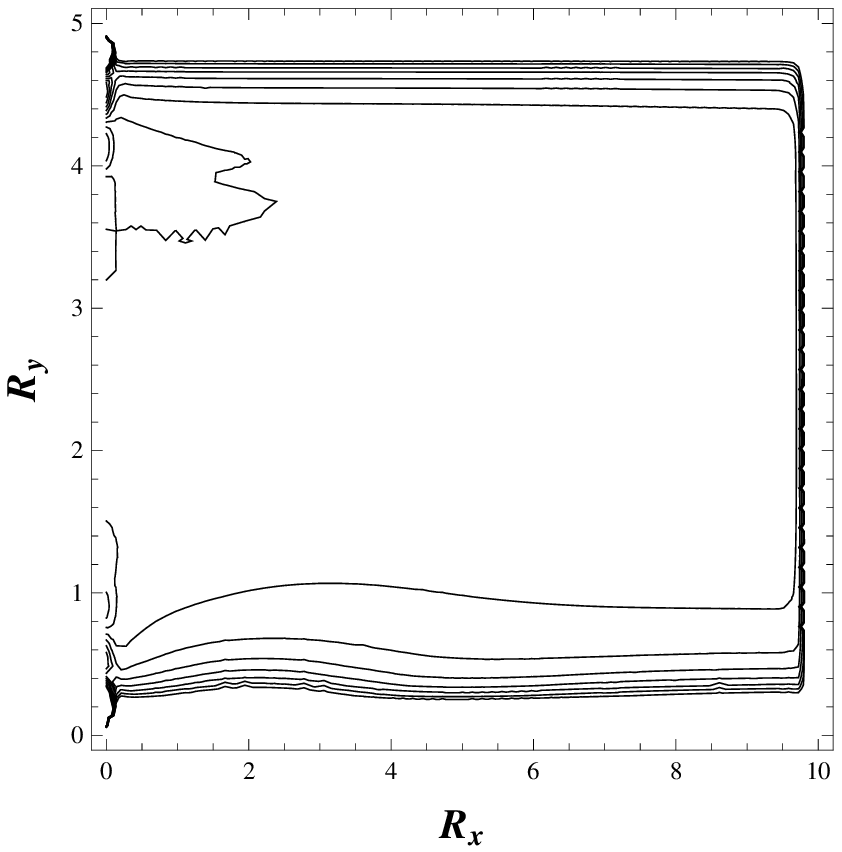}&\includegraphics[width=0.25\textwidth]{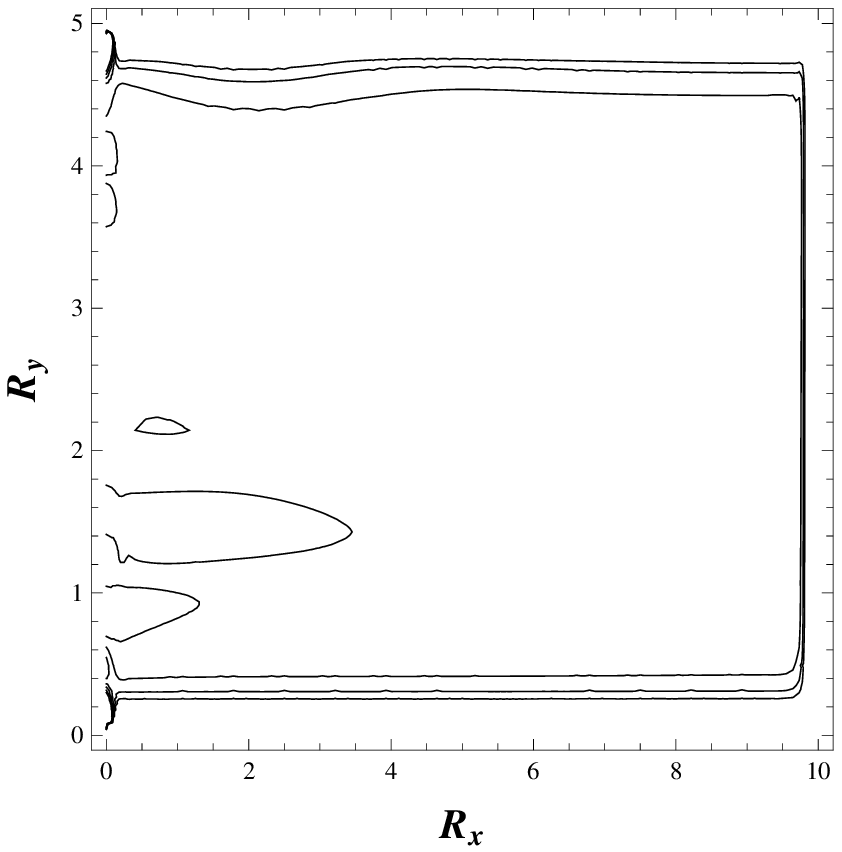} \\
\mbox{(c)}& \mbox{(d)}\\
\end{array}
$
\end{center}
\caption{\label{vdown} (a) The velocity field $\vec{V}=(V_{x},V_{y})$ corresponds to the 
density distribution of Fig. \ref{difu} (a), $\omega>0$. 
(b) The velocity field $\vec{V}=(V_{x},V_{y})$ for $\omega<0$. (c) The corresponding vorticity
 distribution for the vortex configuration of the left panel. The light color indicates 
high vorticity, dark color indicates low vorticity. This figure is a snapshot at time $t=15$ 
for the positive angular frequency case, $\omega>0$. (d) The corresponding 
vorticity distribution $\Gamma_V$ of diffusion equation (\ref{sperm2}) for $\omega<0$.
} 
\vspace{-0.2cm}
\end{figure}

We computed the vortex field of diffusion Eq. (\ref{sperm2}) with respect to the density distribution
 in Fig. \ref{difu} (a). All the parameters have the same value as that of 
Fig. \ref{difu} (a). The vortex configuration appear in the upper half-plane. 
A vortex line(Fig. \ref{vdown} (a)) appears in the density center. 
Noticing the direction of those vectors indicates exactly the direction 
of particle's velocity, the vortex line behaves as a source from which particles are flowing outward. 
The numerical computation of vorticity function(Fig. \ref{vdown} (b)) quantitatively demonstrated 
the distribution of vortex. The high vorticity region are the small inner islands in Fig. \ref{vdown} (b). 
Particles are trapped in the inner island of high vorticity. Further numerical evolution shows, the 
small island of high vorticity is expanding as time increases.

Further more, we computed the force field $\vec{J}=(J_{x},J_{y})$ from density distribution,
\begin{eqnarray}\label{Jxy2}
J_{y}&=&{D_{xy}\omega}\partial_{Rx}\partial_{Ry}{\ln}{N_{s}}+
{D_{xx}\omega}\partial_{Rx}\partial_{Rx}{\ln}{N_{s}},\nonumber\\
J_{x}&=&{D_{yx}\omega}\partial_{Ry}\partial_{Rx}{\ln}{N_{s}}+
{D_{yy}\omega}\partial_{Ry}\partial_{Ry}{\ln}{N_{s}}.
\end{eqnarray}
The corresponding force field $\vec{J}=(J_{x},J_{y})$ with respect to 
the density distribution of Fig. \ref{difu} (a)  is shown in Fig. \ref{difu} (c). 
The direction of force field points to positive X-direction on the upper boundary. 
The force field in the lower half plane points to the upper half-plane. 
In the high density region, the force field shows a turbulence configuration(Fig. \ref{difu} (c)).

%\begin{widetext}

%\end{widetext}

If we switch the angular velocity from a positive $\omega$ to a negative $\omega$, 
both the off-diagonal diffusion tensor and drift tensor will flip a sign simultaneously, i.e., 
\begin{equation}
D_{yx}\Leftrightarrow{-D_{yx}}, \;\;\sigma_{yx}\Leftrightarrow{-\sigma_{yx}}.
\end{equation} 
The off-diagonal diffusion coefficient of cross diffusion term will also flip a sign,
\begin{eqnarray}\label{sperm22}
({D_{yx}+D_{xy}})\Rightarrow-({D_{yx}+D_{xy}}),\;\;\;\;b\Rightarrow-b.
\end{eqnarray}
The drifting velocity of the density center will switch to negative Y-direction. 
The diffusion in X-direction is invariant. The density distribution at time $t=15$ is shown in Fig. \ref{difu} (b) 
for $({D_{yx}+D_{xy}})=1$, b=-5, a=15. The density center moves to the lower half-plane of Y-direction. 
The corresponding velocity field(Fig. \ref{vdown} (a)) with respect to the density distribution of Fig. \ref{difu} (b) 
suggested that the vortex line also shifted to the lower half plane. 
The small inner island of high vorticity distribution quantitatively illustrated the vortex distribution(Fig. \ref{vdown} (d)). 
We computed the effective force field $\vec{J}=(J_{x},J_{y})$. Most force vectors point 
down to the lower half plane(Fig. \ref{difu} (d)) except the turbulent region where the vectors in different 
directions are entangled together.

Numerical computation of the evolution equation of density distribution verified our previous theoretical conclusion.  
When many circling particles are diffusing across a rectangular channel along $X$-axis, 
circles with positive angular frequency drift to the upper boundary(positive Y-direction), 
while circles with negative angular frequency drift to bottom boundary(negative Y-direction).

\subsection{The numerical evolution of density distribution with rotational symmetry}

\begin{figure}
\begin{center}
\includegraphics[width=0.45\textwidth]{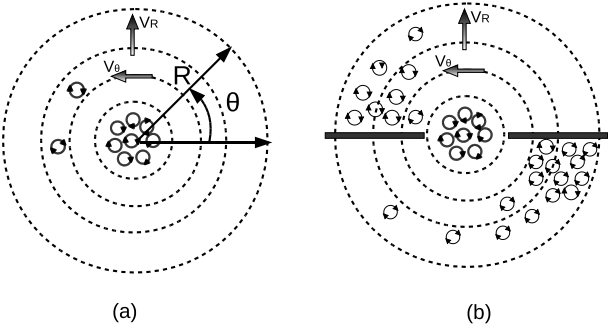}
\caption{\label{cylin} (a) The diffusion of circling particles across a cylindrical density gradient. 
(b) The diffusion phenomena with two blocking walls situated at $\theta=0$ and  $\theta=\pi$.}
\end{center}
\vspace{-0.4cm}
\end{figure}

Confining the particle within a small disc region is more convenient for experimental setup. In that case, 
density distribution has rotational symmetry. The microscopic physics in cylindrical coordinates 
is the same as that in Descartes coordinates. 
We study the special case of constant drift tensor and diffusion tensor. The external force in $\theta$-direction is zero. 
A linearly increasing force is applied along the radius direction,  
\begin{equation}
F_{\theta}=0,\;\;\;F_{R}=C R. 
\end{equation}
The diffusion equation (\ref{polar}) now becomes
\begin{eqnarray}\label{2polar2}
\frac{\partial{N}_{s}}{\partial{t}}&=&{D_{{R}{R}}}\frac{\partial^2{N_{s}}}{\partial{R}^2}
+{D_{{\theta}{\theta}}}\frac{1}{R^2}\frac{\partial^2{N_{s}}}{\partial{\theta}^2}\nonumber\\
&+&[D_{{R}{\theta}}+D_{{\theta}{R}}]\frac{1}{R}\frac{\partial^2N_{s}}{\partial{\theta}\partial{R}}
-D_{{R}{\theta}}\frac{1}{R^2}\frac{\partial{N_{s}}}{\partial{\theta}}\nonumber\\
&-&\sigma_{{R}{R}}C\frac{\partial{N_{s}}}{\partial{\theta}}
-\sigma_{{\theta}{R}}CR\frac{\partial{N_{s}}}{\partial{R}}
-\sigma_{{\theta}{R}}C{N_{s}}.
\end{eqnarray}

\begin{figure}[htbp]
\centering
\par
\hspace{-0.9cm}
\begin{center}\hspace{-0.2cm}
\includegraphics[width=0.3\textwidth]{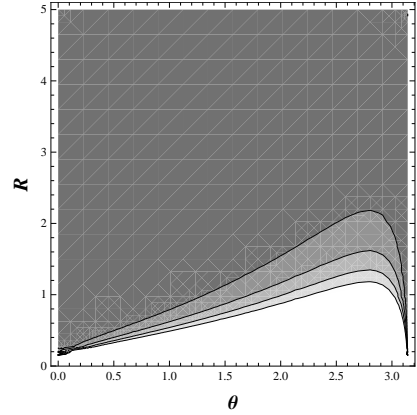}
\end{center}
\caption{\label{po0} The density distribution in $R-\theta$ plane at time 
$t=9$ for $F_{{\theta}}=0, F_{{R}}=0,$ i.e., there is no external force.} \vspace{-0.2cm}
\end{figure}

We first computed the case without any external force, $F_{{\theta}}=0, F_{{R}}=0.$ 
The trajectory of circling particles during diffusion is not only along the radius 
direction, but also has transversal shifting perpendicular to the radius. 
The parameter 
setting for numerical computation are the following: The friction parameter is $\eta=1$. 
The normalization factor $mA^2=1$. Angular frequency is $\omega=10$. The radial diffusion 
constant is $D_{R}=1$. angular diffusion coefficient $D_{\theta}=0.9$. 
The numerical value of diffusion tensors are $D_{{R}{R}}=1$,${D_{{\theta}{\theta}}}=0.9$, 
$D_{{R}{\theta}}+D_{{\theta}{R}}=-1$, $D_{{R}{\theta}}=9$. 
The drift tensor are $\sigma_{{\theta}{\theta}}=\sigma_{RR}=1$, 
$\sigma_{{R}{\theta}}=-\sigma_{{\theta}{R}}=10$. 
In the beginning, the particles concentrate on a small ring with radius 
$R=0.1$ from $0$ to $\pi$. The initial density within the confined region is $1$. 
The numerical evolution of the density distribution at time $t=9$ is shown in Fig. \ref{po0}. 
The particles concentrating around $\theta=\pi$ is much more than that around $\theta=0$.

\begin{figure}[htbp]
\centering
\par
\begin{center}
$
\begin{array}{c@{\hspace{0.03in}}c}
%\multicolumn{1}{1}{\mbox{}} & \multicolumn{1}{1}{\mbox{}} \\
\hspace{-0.2cm}
\includegraphics[width=0.24\textwidth]{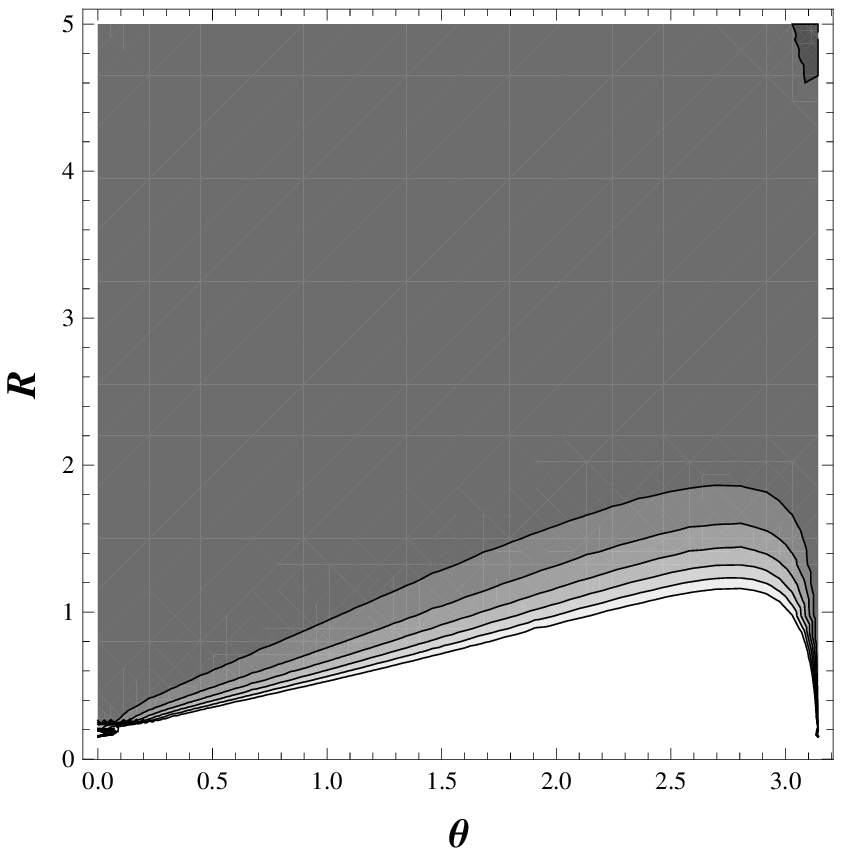}& \includegraphics[width=0.24\textwidth]{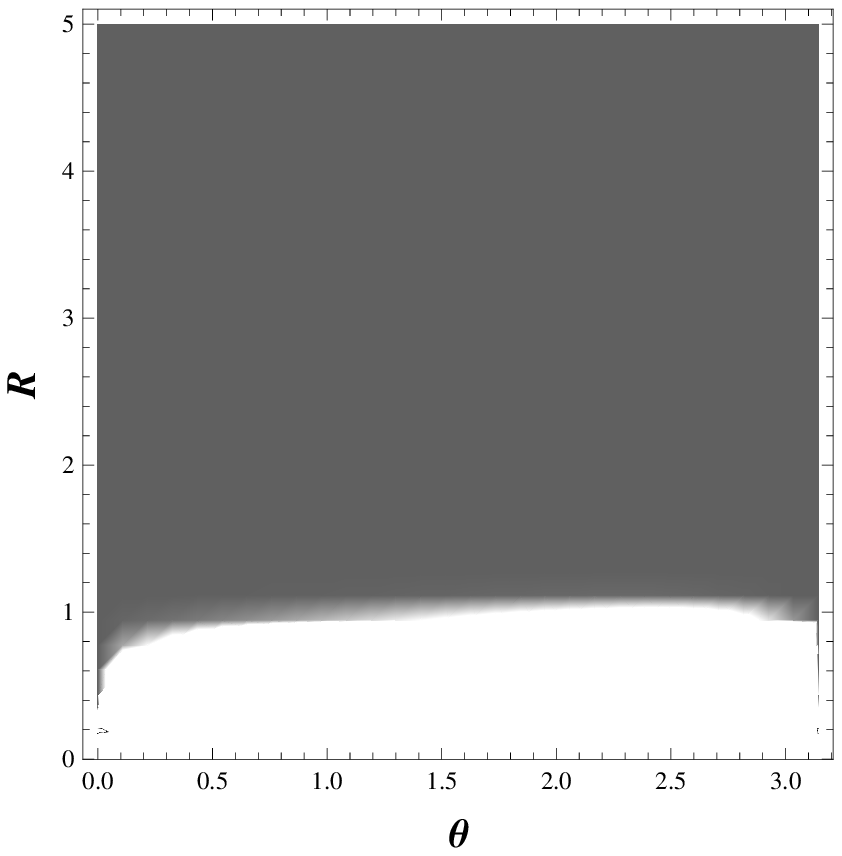}\\
\mbox{(a)} & \mbox{(b)}\\
\end{array}
$
\end{center}
\caption{\label{po182}(a) The density distribution in $R-\theta$ plane at time $t=9$. 
The slop of the external force is $C=1/8$.(b) The density distribution in $R-\theta$ 
plane at time $t=9$. The slop of the external force is $C=2$..} \vspace{-0.2cm}
\end{figure}

When we increase the external field along the radius, this bias concentration phenomena will be suppressed. 
We numerically computed the density distribution for $F_{{\theta}}=0, F_{{R}}=R/8 $, 
all the other parameters have the same value as that for $C=0$. 
The amplitude of the density center near $\pi$ is shrinking(Fig. \ref{po182} (a)). 
For $F_{{\theta}}=0, F_{{R}}=2 R$, the density distributed in the block zone 
$(0,\pi)$ becomes almost homogeneous(Fig. \ref{po182} (b)). 
Thus if the external force along radius is too strong,  an apparent bias density distribution 
is hardly observable because of the dominant strong flow along radius, the transverse flow in $\theta$-direction 
is ignorable.

We propose an experiment to illustrate the transverse flow of circling particles diffusing 
in a disk region. First, we confine many circling particles in a very small disk zone. 
Second, we place two blocking wall along the radius, one is along $\theta=0$, the other 
is along $\theta=\pi$. Finally we withdraw the confining potential around the small disk to 
let them diffuse.

If there is no blocking walls, the density distribution expands circles around the origin. 
The transverse flow $\vec{V}_\theta$ perpendicular to the radius can not be observed 
from density distribution, for the particle flow has the same rotational symmetry as density distribution. 
The blocking wall break the radial symmetry by cutting the transverse flow into two separated parts. 
Since the flow has only one direction, the particles will concentrate only on one side of wall. 
Only a few particles appear on the other side of the same wall. 
Fig. \ref{cylin} (b) showed the accumulation phenomena for particle circling in clockwise direction.
 If the particle is circling in the counterclockwise direction, the particle will 
accumulate in the opposite side of the wall. 
If there exist two kinds of particles, one is circling in clockwise direction, 
the other is circling in counter clock wise direction. 
The particle circling in clockwise direction will accumulate in the region $\theta=(-\alpha,0)$ and $\theta=(\pi-\alpha, \pi)$, 
$\alpha>0$. The particle circling in counterclockwise direction will accumulate on 
the opposite side of the wall, i.e., $\theta=(0,\beta)$ and $\theta=(\pi, \pi+\beta)$, $\beta>0$. 
This provides us one way to divide particles with opposite circling direction.

\section{Theoretical understanding on the diffusion experiment of many sperms}

We first introduce some background knowledge of sperm for understanding the many sperm diffusion experiment\cite{inamdar}.  
The average size of an ordinary sperm head is about 5 $\mu{m}$. Its flagellum has a length of 30-60 $\mu{m}$. 
In three dimensions, the trajectory of a swimming sperm is a helical curve. 
If there is no external stimulus, the center-line of the helical curve is almost 
straight\cite{grenshaw}. When chemoattractant spreads into the solution, 
the sperm will reorient its swimming direction under stimulus, and. 
It is believed that the sperm will bent their circling direction to the 
source of chemoattractant\cite{julicherPNAS}\cite{grenshaw}. 
A sperm swims in circles near a boundary. The chemical stimulus of chemoattractant acts as 
some effective attractive action on sperms 
which change the velocity of sperm and drive sperm to eggs\cite{Bohmer}\cite{eisenbach}.

In the many sperm diffusion experiment\cite{inamdar}, the sperms are kept in a sperm reservoir in 
the beginning. The sperm reservoir is placed at one end of a rectangular migration channel. On the 
other end of the rectangular migration channel is a chemoattractant reservoir. 
The sperm used in experiment\cite{inamdar} is \emph{Arbacia punctulata} spermatozoa. 
The image of density distribution in experiment\cite{inamdar} shows that sperm liken to concentrate 
on upper boundary of the channel(Fig.\ref{expe}), and the sperm density concentration 
illustrated a plateau along the length of channel. The position of the plateau 
is near the reservoir of sperm(around $x_2$ in Fig. \ref{expe})\cite{inamdar}.

\begin{figure}[htbp]
\centering
\par
\hspace{-0.9cm}
\begin{center}\hspace{-0.2cm}
\includegraphics[width=0.4\textwidth]{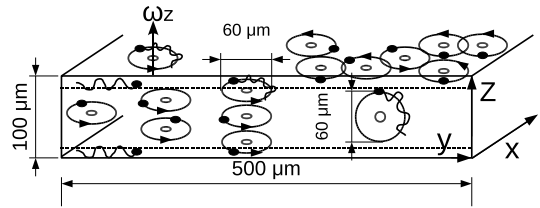}
\end{center}\vspace{-0.4cm}
\caption{\label{s} A cross section of the migration channel of experiment\cite{inamdar}. 
The length along $X$-axes is 7000-9000 $\mu$m. The width in $Y$-axes is 500 $\mu$m, and the height in $Z$-axes is 100 $\mu{m}$. The radius of the sperm circle is $30$ $\mu$m.} \vspace{-0.2cm}
\end{figure}

The rectangular migration channel in experiment\cite{inamdar} has a length of 7000-9000 $\mu$m, 
its width is 500 $\mu$m, and the height of the channel is 100 $\mu{m}$. 
This channel is a three dimensional channel, but for many sperms, 
it is well approximated by a two dimensional channel. 
The radius of the sperm circle is fluctuating around $30$ $\mu{m}$\cite{riedel}\cite{woolley}\cite{Bohmer}\cite{eisenbach}. 
So one sperm need at least a small square of $60\times60$ $\mu{m}^2$ to swim. 
The height of channel is 100 $\mu{m}$ which only allows one sperm circle to rotate freely in three dimensions. 
The angular velocity of one sperm can rotate from ${\omega}_z$ to ${\omega}_x$ or to ${\omega}_y$. 
But when we consider two sperms, one is on top of another along $Z$-direction. 
The height of the channel must be at least $120$ $\mu{m}$ so that the two sperms can simultaneously
rotate their angular velocity without any conflicting. For a channel with a height 100 $\mu{m}$, 
the freedom of angular velocity will be greatly limited for many sperms. 
Following the same analyze, three sperms on top of one another along $Z$-direction needs at least a channel with height $180$ $\mu{m}$ 
rotates simultaneously without any conflicting.

The sperm experiment\cite{inamdar} study the diffusion of millions of sperms. 
The sperms in the top surface of the channel will swim in circles. The thickness of the top layer is about 5 $\mu{m}$ 
which is the size of one sperm head. The top layer behaves as the boundary for the sperms below, 
the sperms in the second layer will also swim in circles, and so does the third layer, and so on. 
Many sperm system of high density can be approximated by a two dimensional system. 
The best way of creating a good two dimensional many sperm system is to 
reduce the height of the rectangular channel in experiment\cite{inamdar} 
from $100$ $\mu{m}$ to $5\sim10$ $\mu{m}$. In that case, the angular velocity in X and Y direction, 
${\omega}_x,{\omega}_y$, will be greatly suppressed. The dominant component of angular velocity 
is ${\omega}_z$. The height of channel is only free enough to hold one layer of sperms, 
thus it is effectively two dimensional system.

\begin{figure}[htbp]
\centering
\par
\begin{center}
$
\begin{array}{c@{\hspace{0.03in}}c}
%\multicolumn{1}{1}{\mbox{}} & \multicolumn{1}{1}{\mbox{}} \\
\includegraphics[width=0.3\textwidth]{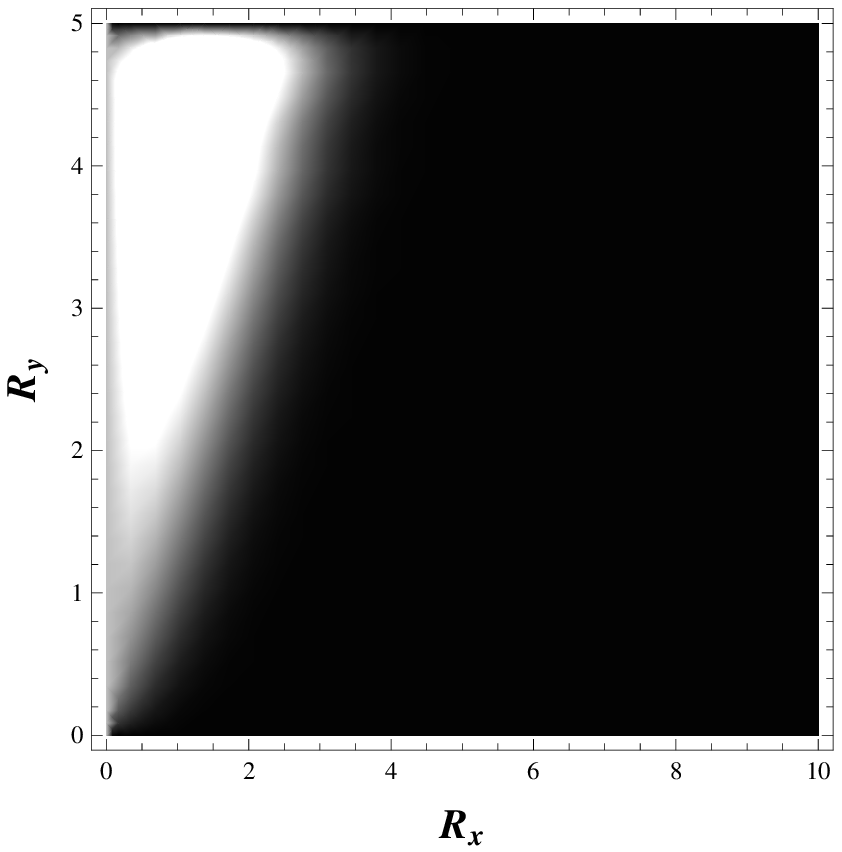}& \\
\mbox{(a)} &\\
\includegraphics[width=0.3\textwidth]{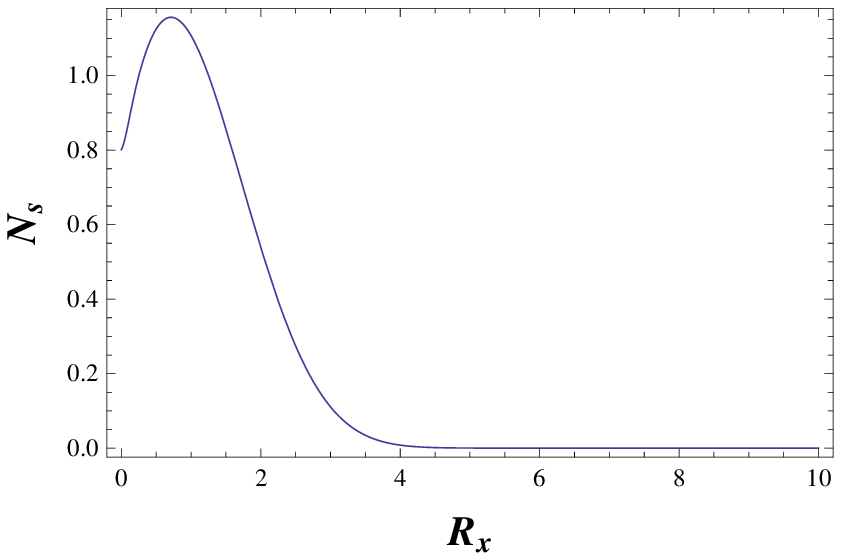}&\\
\mbox{(b)} & \\
\end{array}
$
\end{center}
\caption{\label{expd2} (a)The density distribution $N_{s}(R_x,R_y,t=15)$ of diffusion 
Eq. (\ref{finaldifu}) for $C=2$. The light color indicates high density, dark color indicates low density.
(b) The density distribution of $N_{s}(R_x,t=15)=\int_0^{h}N_{s}dR_y$ 
for diffusion Eq. (\ref{finaldifu}) at $C=2$.} \vspace{-0.2cm}
\end{figure}

We apply the two dimensional evolution equation of density distribution as an approximation 
of the many sperm diffusion experiment\cite{inamdar}. There is a chemo-reservior attached on the other 
end of the channel, we assume the chemoattrative force distribution is
\begin{equation}
F_{x}=C\;R_x, \;\;\;F_{y}=0. 
\end{equation}
This distribution means a sperm feels stronger chemoattractive force if it is closer to the chemoattractant. 
Biologist found that almost all sperms are circling with same handedness\cite{riedel}. 
So we choose angular velocity $\omega_z>0$. The diffusion equation for this external driven force is
\begin{eqnarray}\label{finaldifu}
\frac{\partial{N}_{s}}{\partial{t}}&=&D_{xx}\partial_{Rx}^{2}N_{s}+D_{yy}\partial_{Ry}^{2}N_{s}\nonumber\\
&+&({D_{yx}+D_{xy}})\partial_{Rx}\partial_{Ry}N_{s} \nonumber\\
&-&C\;\sigma_{yx}\;R_x\;{\partial_{Ry}{N_{s}}}-C\;\sigma_{xx}\;R_x\;{\partial_{Rx}{N_{s}}}\nonumber\\
&-&C\;\sigma_{xx}{N_{s}}.
\end{eqnarray}
We take the diagonal diffusion coefficient as $D_{xx}=0.9,\;D_{yy}=1$. The off-diagonal 
diffusion coefficient is ${D_{yx}+D_{xy}}=-1$. 
The off-diagonal drifting tensor is $\sigma_{yx}=5$. 
The diagonal drifting tensor is $\sigma_{xx}=2$. 
The slop of the external force is controlled by $C$. 
If we change $C$, three terms in Eq. \ref{finaldifu} will change simultaneously. 
We first study the case of $C=2$. 
The numerical evolution of density distribution at 
time $t=15$ is shown in Fig. \ref{expd2} (a)). 
The density center shift to positive Y-direction. 
We integrate the density along Y-axis $N_{s}(x,t)=\int_0^{h}N_{s}dy$, 
and plot the density evolution $N_{s}(x,t)$ along $X$-direction
(Fig. \ref{expd2} (b)). The center of the density distribution is 
around $R_x=2$. The diffusion curve along $R_X$ is not exponential decay. 
The density center propagate as a pump.

\begin{figure}[htbp]
\centering 
\par
\begin{center}
$
\begin{array}{c@{\hspace{0.03in}}c}
%\multicolumn{1}{1}{\mbox{}} & \multicolumn{1}{1}{\mbox{}} \\
\vspace{-0.1cm}\hspace{-0.4cm}
\includegraphics[width=0.245\textwidth]{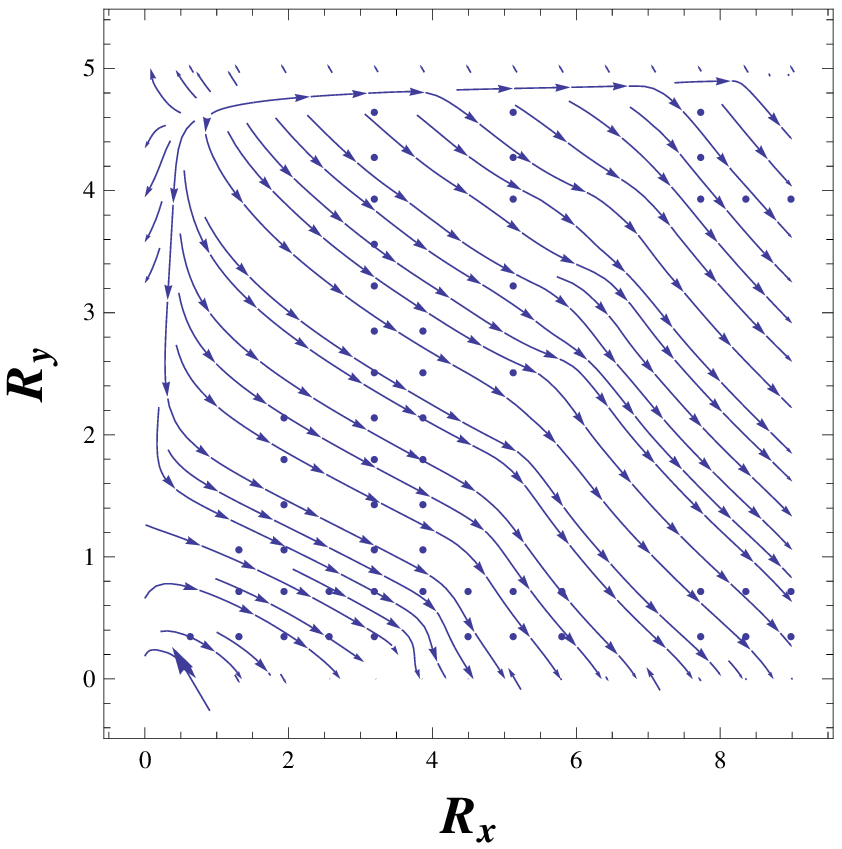}& \includegraphics[width=0.245\textwidth]{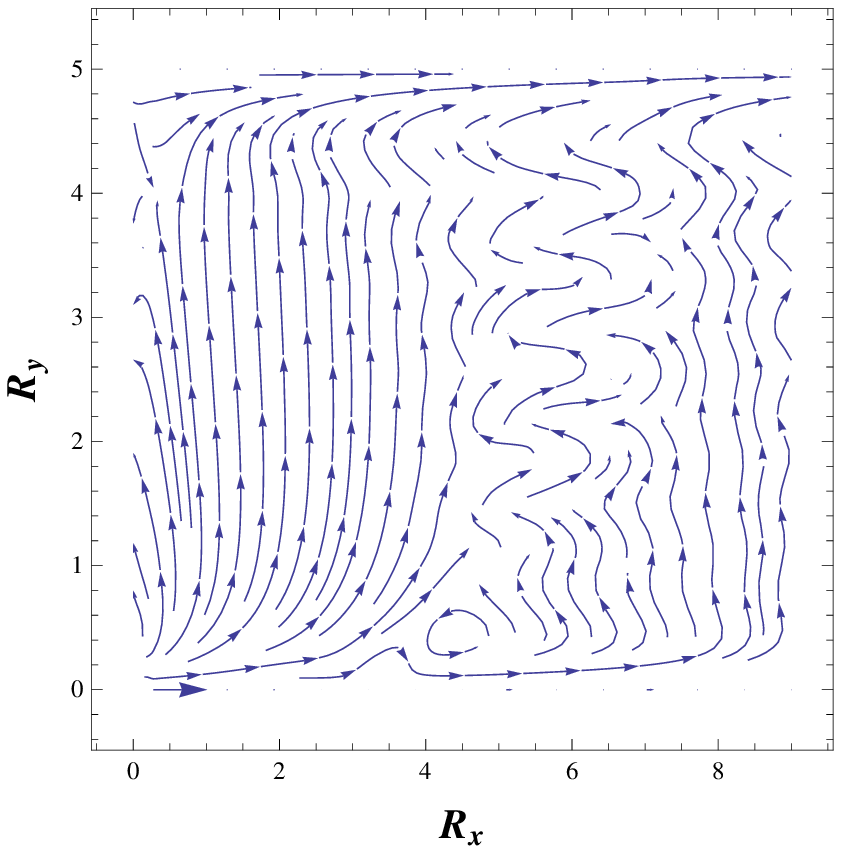}\\
\mbox{(a)} & \mbox{(b)}\\
\end{array}
$
\end{center}
\caption{\label{expv2} (a) The velocity field distribution $\vec{V}=({V}_x,{V}_y)$ 
corresponding to the density distribution Fig. \ref{expd2}. The slop of the external force is $C=2$. .
 (b) The effective force field $\vec{J}=({J}_x,{J}_y)$ at 
time $t=15$ corresponding to the density distribution Fig. \ref{expd2}. The 
slop of the external force is $C=2$.} \vspace{-0.2cm}
\end{figure}

The vortex generated during diffusion will reduce the speed of diffusion in $X$-direction. 
The vortex core will attract many particles to flow around the core. The velocity of particle is given by 
\begin{eqnarray}
\dot{R}_x&=&-{D_{xy}}\frac{\partial_{Ry}{N_{s}}}{N_{s}}-
{D_{xx}}\frac{\partial_{Rx}{N_{s}}}{N_{s}}+\sigma_{xx}C\;R_x,\nonumber\\
\dot{R}_y&=&-{D_{yx}}\frac{\partial_{Rx}{N_{s}}}{N_{s}}
-{D_{yy}}\frac{\partial_{Ry}{N_{s}}}{N_{s}}+\sigma_{yx}C\;R_x.
\end{eqnarray}
We computed the velocity field distribution at 
time $t=15$ in $R_x-R_y$ plane, 
a vortex line appears in the high density region(Fig. \ref{expv2} (a)). 
Particles are flowing out of these vortex lines.

We also computed the corresponding force field $\vec{J}$ induced by velocity gradient,
\begin{eqnarray}
J_{y}&=&{D_{xy}\omega}\partial_{Rx}{\partial_{Ry}{\ln}{N_{s}}}+
{D_{xx}\omega}\partial_{Rx}{\partial_{Rx}{\ln}{N_{s}}}\nonumber\\
&+&\omega\sigma_{xx} C\nonumber\\
J_{x}&=&{D_{yx}\omega}\partial_{Ry}{\partial_{Rx}{\ln}{N_{s}}}+
{D_{yy}\omega}\partial_{Ry}{\partial_{Ry}{{\ln}N_{s}}}.
\end{eqnarray}
The strong flow of $\vec{J}$ is distributed in the high density region, 
most vectors points to positive $R_y$ direction(Fig. \ref{expv2} (b)). On the upper boundary, 
all flows point to positive $R_x$ direction. 
Experiment observed that sperms on the upper 
boundary swim faster than others. The force field on the upper boundary is stronger than other region.
This provides a possible understanding to the fast sperms on the upper boundary.

\begin{figure}[htbp]
\centering
\par
\begin{center}
$
\begin{array}{c@{\hspace{0.03in}}c}
%\multicolumn{1}{1}{\mbox{}} & \multicolumn{1}{1}{\mbox{}} \\
\includegraphics[width=0.3\textwidth]{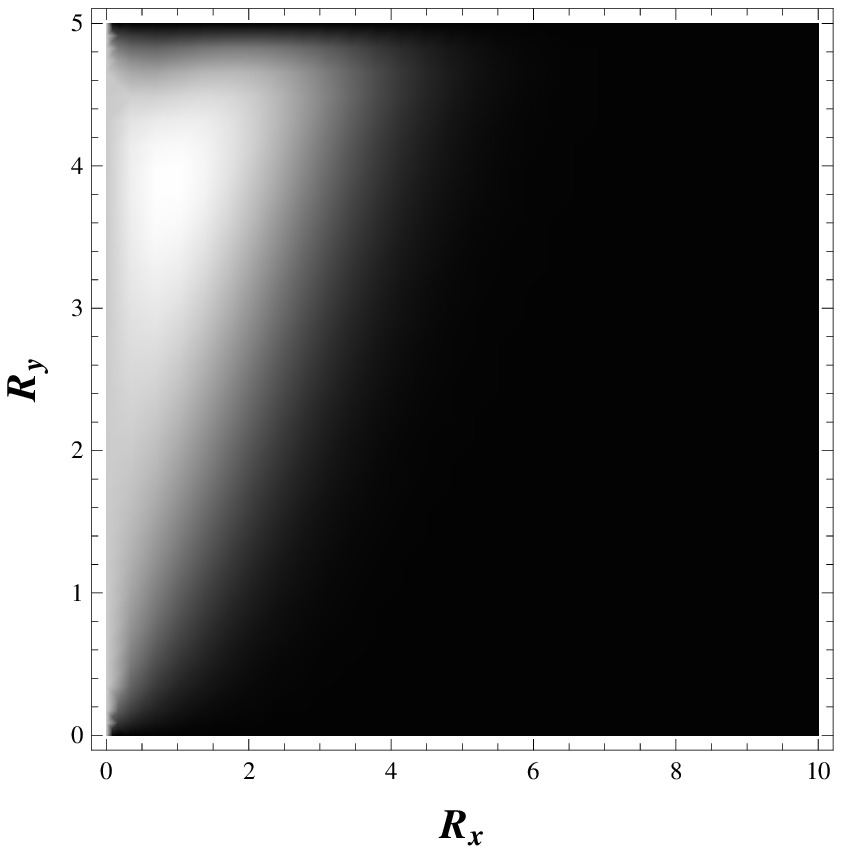}& \\
\mbox{(a)} &\\
\includegraphics[width=0.3\textwidth]{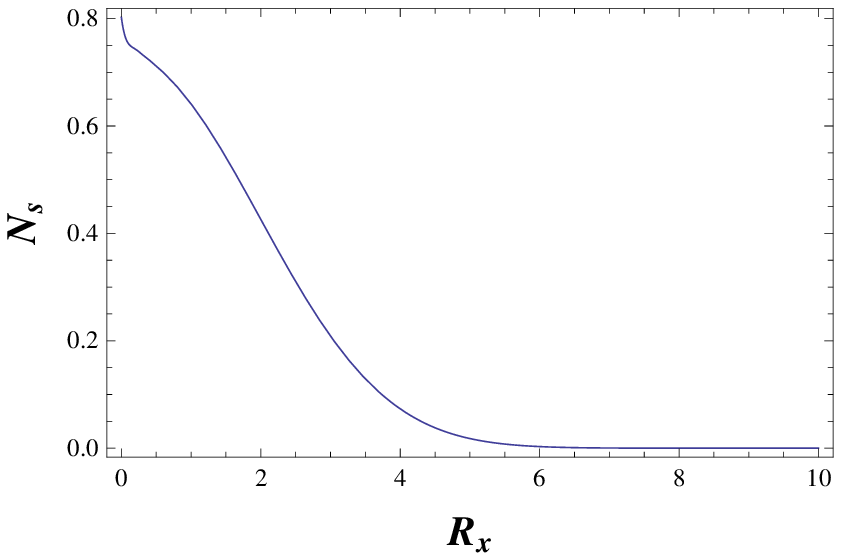}&\\
\mbox{(b)} & \\
\end{array}
$
\end{center}
\caption{\label{n4} (a)The density distribution $N_{s}(R_x,R_y,t=15)$ of diffusion Eq. (\ref{finaldifu}) for $C=1/2$. 
The light color indicates high density, dark color indicates low density. 
(b) The density curve of $N_{s}(R_x,t=15)=\int_0^{h}N_{s}dR_y$ for $C=1/2$.} \vspace{-0.2cm}
\end{figure}

The experiment \cite{inamdar} measured the density distribution from one snapshot of diffusion movie. 
The experiment date of Ref. \cite{inamdar} is collected from the diffusion image 
along the length of the migration channel. 
A point on the curve represents the average density of sperms in each 
small rectangular box along Y-axis(the gray box in Fig. \ref{expe}). 
Ref. \cite{inamdar} fitted the data by one dimensional diffusion equation. 
However there is apparent discrepancy between experiment data and the theoretical curve of one dimensional theory. 
They measured the diffusion of \emph{Arbacia $\;$punctulata} spermatozoa in 
different solutions. In all of these measurements, the experiment data near high density region
deviates apparently from exponential decay around $x=2$ in Fig. \ref{expe}.
 I believe these discrepancy is not accident or erroneously 
fluctuating from experimental environment. I performed numerical evolution of the two dimensional diffusion equation. 
The maximal peak of density distribution around $x=2$ is propagating along the channel. 
I compared the density distribution for different slope of external force 
after the same time interval, $C=2$(Fig. \ref{expd2} (b)) 
and  $C=1/2$(Fig. \ref{n4} (b)). The density center of $C=2$ travels faster than that of $C=1/2$.

The same data of density distribution in Ref.\cite{inamdar} also demonstrated a static peak siting right 
at the origin $R_x=0$. This highest density peak at the origin preserves in different experiment data of one 
dimensional density distribution. I guess the appearance 
of this high density peak is due to experiment devices. The sperm reservoir is right at the origin. 
Biological observation suggest that sperms like to stay near a boundary. 
If the sperms are attached by the wall of the channel in the beginning, 
they would not move too much in the whole period of diffusion. 
The peak at the origin probably has no dynamic effect. 
The data in Ref. \cite{inamdar} is obtained from only one snapshot of the whole diffusion 
movie. We do not have enough data to draw a definite conclusion.

\section{Conclusion}

Alive circle swimmers, such as bacterium and sperms, are macroscopic objects comparing with the scale of water molecules. 
Usually the interaction between the beating flagella of the swimmer and the thick fluidics environment 
provide the driving-force for swimming. The beating modes of flagella may depend on internal state of the swimmer. 
One can construct a biophysical model of alive circle swimmer by coupling Newton's law of motion 
and the internal state equation of the biological or chemical process inside the swimmer. 
In this article, we only focus on the general physics of Newtonian equation of 
motion for circling particles. The force in Newtonian equation is a general function 
which can be interpreted as state-dependent force. How to control this state-dependent force is 
an independent issue of biology or chemistry.

We take abstract point particles as analogy of those tiny microorganism. The ideal point particle are circling 
around a drifting center. The circling speed is much larger than the drifting speed of the circle center. 
The point particle is moving in an environment with high frictional resistant force. 
The local circling motion is fast enough to counterbalance the frictional force. 
Local circling motion is not overdamped. The drifting motion of the center is overdamped 
for the slowly drifting center is too weak to cancel the frictional force. 
We first established the dynamic equations of a single circling particle and the corresponding probability conservation equation, 
and then developed the evolution equation of density distribution of many circling particles from 
the particle number conservation law.

Comparing with non-circling particle, the most special character of circling particle is that its 
velocity in one directions is strongly coupled to the velocity component in other direction. 
This strong correlation comes from the geometric constrain of a closed circular trajectory. 
In fact, this geometric constrain is a special case of topological constrain. As long as 
the local trajectory of a moving particle is exactly closed, the strong correlation between 
different velocity components will exist. The trajectory could be a deformed circle or ellipse, 
or other closed curve which is homomorphic to a circle. We can define a mathematical 
centripetal force from an arbitrary mathematical trajectory.

We first established the equation of motion for a single circling particle from Newton's law of 
motion. A non-zero external force in one direction will drive the circle center 
moving to the direction perpendicular to this direction. 
This theoretical result can help us to understand a recent numerical simulation of the motion a single sperm under 
complex chemical stimulus\cite{julicherPNAS}. If the external force does not exist, 
the circle center of single circling particle will just randomly drift around in a thermal environment.

There exist intrinsic difference between the diffusion of many circling particles 
and the diffusion of probability distribution of single circling particle. 
Density gradient only exist for many particle system. It is due to the density gradient that 
one circling particle can feel the existence of other circling particles. A single circling 
particle must have long memory to have similar physics even though 
it is not exactly the same physics. We established the evolution equation 
of density distribution for many circling particles from Newton's law of motion 
by imposing a conservation equation of total particle number. 
When there is no any external force, the particles flowing along one direction 
of the density gradient can induce a flow perpendicular to this direction. 
This phenomena does not occur during the evolution of probability 
distribution of single circling particle without external force.

Recently a diffusion experiment of many circling sperms had been conducted in a rectangular channel\cite{inamdar}. 
We performed numerical evolution of the density distribution in a rectangular diffusion channel to meet 
the experimental setup. When we confine the particles to one end of the channel and then release them to diffuse 
along the channel, the particles circling in 
clockwise direction concentrates on one side of channel, particles circling in counterclockwise 
direction concentrates on the other side of channel. One of the experiment observations\cite{inamdar} 
indeed showed apparent bias density concentration which supports our theoretical prediction. 
The density distribution of other images in experiment Ref. \cite{inamdar} is not obvious. According to 
our numerical computation of density distribution for different external forces, if the external force is too strong, 
the phenomena of bias density distribution will be suppressed. The qualitative image of camera is not 
accurate enough to capture the fine structure of density distribution. We computed the quantitative density 
distribution along the channel. The numerical evolution of the density distribution 
looks very similar to the density curve of experiment data. Therefore one possible theoretical explanation 
for those images without obvious bias density concentration is that 
the effective attractive force of chemoattractant is too strong.

The sperm experiment in Ref. \cite{inamdar} also observed that sperms swim faster on the boundary. 
We computed the velocity vector field and the force field. Most vectors in the force field point to the boundary. 
The strongest force field distribute right along the boundary. This probably can help us understand why sperms 
on the boundary swim faster. More over, vortex are also generated during diffusion. Numerical computation shows, 
the high vorticity region is near the boundary. The trapping of particles in the vortex will increase the local 
density which in turn increase the density gradient force. Vortex maybe is another reason of faster sperms 
on the boundary.

So far, my theory agrees with most experimental observations. Further more, 
we predict one phenomena: if one mix the circling particles with opposite circling handedness 
together, and let them diffuse across a rectangular channel in X-direction. They will split into two 
groups, one goes to positive Y-direction, the other goes to negative Y-direction. The particles in 
each group have opposite circling handedness to the other group.

The collective phenomena of many circling microorganism is far more complicate than classical physics particle. 
Besides various different biological and chemical factors, the physics of many circling particles within viscous 
liquid is not well studied so far. One may develop other theoretical approaches by taking 
into account of more detail biological parameters. For example, the special shape of the circling particle maybe important for swimming in liquid with 
low Reynolds number. The sperm head has a shape of rod. The beating head of a boundary swimmer may hit the wall before its beating 
amplitude reach maximal point. As a swimmer liken to slide along the lengthy 
direction of the head, the cut off of beating amplitude in Y-direction by the wall 
will enhanced its motion in $R_X-$direction. If a rod is surrounded by many rods, 
one may include the collisions between different particles. 
Some well known equations, such as Boltzmann equation, Fokker-Planck equation, 
or fluid dynamics\cite{yang}, or plasma physics\cite{ffchen}, may also shed new light 
on the physics of circling particles. They are more or less related, for example, 
Fokker-Planck equation is linear approximation of Boltzmann equation\cite{pawula}. 
Living microorganism seldom collide with each other by obeying local momentum conservation law exactly. 
They are able to avoid each other. The dynamic equation of a velocity-dependent distribution function 
for living microorganism is beyond the scope of Boltzmann equation.

\section{Acknowledgment}

I thank Dr. Janik Kailasvuori for a critical reading of this manuscript.

\appendix

\section{The velocity and acceleration of a helical trajectory in two dimensions}

The author did not find such a calculation in textbooks or references that follows similar logic chain as I did here. 
So I add the detail calculation in this section to show centripetal force is a mathematical
 conception that can be deduced from an arbitrary helical curve.

If the center of the circle is not fixed and drifts around at low speed, the trajectory of a circling particle 
is a helical curve, it can be described by drifting circles. We introduce vector, 
$\vec{R}(t)=R_{x}(t)\textbf{e}_{x}+R_{y}(t)\textbf{e}_{y}$, 
to express the instantaneous center of the circle. The relative position 
of the particle with respect to the center is denoted as 
$r(t)=r_{x}(t)\textbf{e}_{x}+r_{y}(t)\textbf{e}_{y}=r_{0}(t)e^{i\omega{t}}$, $r_{0}(t)$ is 
the relative distance between a particle and the center of circle. 
The complete position coordinates of the circle swimmer is 
\begin{eqnarray}
x(t)\textbf{e}_{x}+y(t)\textbf{e}_{y}=\vec{R}(t)+\vec{r}(t)=\vec{q}(t), 
\end{eqnarray}
\begin{eqnarray}
x(t)&=&{R}_{x}(t)+r_{x}(t),\;\;\;\;r_{x}(t)={r}_{0}(t)\cos({\omega} t),\nonumber\\
y(t)&=&{R}_{y}(t)+r_{y}(t),\;\;\;\;r_{y}(t)= {r}_{0}(t)\sin({\omega} t).
\end{eqnarray}
This coordinates provide us a convenient mathematical description for an arbitrary helical curve in two dimensional space.  The shape of the curve depends on the function of center coordinates $R(t)=R_{x}(t)+iR_{y}(t)$ and the function of radius ${r}_{0}(t)$. For example, if the center accelerates in X-direction with a constant acceleration $a$, and accelerates in Y-direction with acceleration $b$,  the center coordinates functions are $R_{x}(t)=\frac{1}{2}a\;t^{2}$, $R_{y}(t)=\frac{1}{2}b\;t^{2}$. In the mean time, if we assume the radius of the circle is fluctuating periodically, ${r}_{0}(t)={r}_{0}+cos(kt)/100$, the equation for a drifting helical curve is 
\begin{eqnarray} 
x(t)&=&\frac{1}{2}a\;t^{2}+\left[{r}_{0}+\frac{\cos(kt)}{100}\right]\cos{\omega{t}},\;\;\;\nonumber\\
y(t)&=&\frac{1}{2}b\;t^{2}+\left[{r}_{0}+\frac{\cos(kt)}{100})\right]\sin{\omega{t}}. 
\end{eqnarray}

From physics point of view, the tangent vector of the curve is the velocity of a particle 
which move along the curve. The acceleration can be expressed as the curvature of the curve. 
The time parameter $t$ can be viewed as an arc length parameter of the curve. The 
first derivative of curve ${q(t)}$ with respect to $t$ give us the velocity,
\begin{eqnarray}
\dot{x}&=&\dot{R}_{x}(t)-v_0(t)\sin({\omega} t)+\dot{r}_{0}(t)\cos({\omega} t),\nonumber\\
\dot{y}&=&\dot{R}_{y}(t)+v_0(t)\cos({\omega} t)+\dot{r}_{0}(t)\sin({\omega} t). 
\end{eqnarray}
$\dot{R}_{x}(t)$ and $\dot{R}_{y}(t)$ are the drifting velocity of the center. $v_0(t)=r_{0}(t){\omega}$ is the speed of circling motion around the center. The mathematical definition of circle swimmer requires that the circling speed of swimmer is much larger than the speed of its drifting circle center, i.e., $v_0(t)\gg\dot{R}_{\alpha}(t), \alpha=x,y$. For the opposite case, $v_0(t)\ll\dot{R}_{\alpha}(t)$, the trajectory of the particle is no longer drifting circles, and becomes an arbitrary curve. In that case, one can choose differential geometry to construct an equivalent theoretical description. The second derivative of ${q(t)}$ with respect to $t$ reads
\begin{eqnarray}\label{acxy}
\ddot{x}&=&\ddot{R}_{x}(t)-\dot{y}{\omega}+{\omega}\dot{R}_{y}(t)+\ddot{r}_{0}(t)\cos({\omega} t)\nonumber\\
&-&\dot{r}_{0}(t){\omega}\sin({\omega} t),\;\nonumber\\
\ddot{y}&=&\ddot{R}_{y}(t)+\dot{x}{\omega}-{\omega}\dot{R}_{x}(t)+\ddot{r}_{0}(t)\sin({\omega} t)\nonumber\\
&+&\dot{r}_{0}(t){\omega}\cos({\omega} t).
\end{eqnarray}
The acceleration has two parts: the acceleration of the center, the acceleration of circling motion around the center,
\begin{eqnarray}\label{acxy2}
\ddot{x}&=&[\ddot{R}_{x}(t)+\ddot{r}_{0}(t)\cos({\omega} t)]\nonumber\\
&-&[\dot{y}{\omega}-{\omega}\dot{R}_{y}(t)+\dot{r}_{0}(t){\omega}\sin({\omega} t)],\;\nonumber\\
\ddot{y}&=&[\ddot{R}_{y}(t)+\ddot{r}_{0}(t)\sin({\omega} t)]\nonumber\\
&+&[\dot{x}{\omega}-{\omega}\dot{R}_{x}(t)+\dot{r}_{0}(t){\omega}\cos({\omega} t)].
\end{eqnarray}
We summarize the position of the particle as one vector
\begin{eqnarray}
x(t)\textbf{e}_{x}+y(t)\textbf{e}_{y}=\vec{R}(t)+\vec{r}(t)=\vec{q}(t), 
\end{eqnarray}
Then the acceleration has more compact formulation,
\begin{eqnarray}\label{a=r+R2}
\ddot{\vec{q}}(t)=\vec{a}-{\vec{V}}\times\vec{\omega},\;\;\;\; \vec{\omega}=\omega{\textbf{e}_{z}},
\end{eqnarray}
where $\vec{a}$ is the absolute acceleration of the circle center, 
\begin{eqnarray}\label{ac}
a_{x}&=&[\ddot{R}_{x}(t)+\ddot{r}_{0}(t)\cos({\omega} t)], \;\;\nonumber\\
a_{y}&=&[\ddot{R}_{y}(t)+\ddot{r}_{0}(t)\sin({\omega} t)]. 
\end{eqnarray}
$\vec{V}$ is the relative velocity with respect to the center,
\begin{eqnarray}\label{vc}
V_{x}&=&[\dot{x}-\dot{R}_{x}(t)+\dot{r}_{0}(t)\cos({\omega} t)].\nonumber\\
V_{y}&=&[\dot{y}-\dot{R}_{y}(t)+\dot{r}_{0}(t)\sin({\omega} t)].
\end{eqnarray}
So far, the acceleration and velocity mentioned above are completely mathematical conception defined along a trajectory curve. 
One can rewrite everything above in terms of differential geometry. It is only when we apply Newton's law, 
$F=m\ddot{\vec{q}}(t)$, the geometric quantity above is related to physical force, i.e., 
\begin{eqnarray}
\vec{F}=m\ddot{\vec{q}}(t)=m\vec{a}-m{\vec{V}}\times\vec{\omega},
\end{eqnarray}
where the first term $F_{external}=m\vec{a}$, 
and the second term is $ F_{centripetal}=-m {\vec{V}}\times\vec{\omega}$. 
The non-zero angular frequency $\omega$ summarized the effective centripetal force for the circling motion. 
Eq. (\ref{vc}) tells us the centripetal force $m {\vec{V}}\times\vec{\omega}$ is consumed by three motions. 
One is circling motion around the center. The second consumer is the motion of the center. The third consumer 
is the expanding or shrinking of the radius. The other external effective driven force $F_{external}$ determines 
the acceleration of the circle center $m\vec{a}$. The angular frequency $\omega$ may have different physical origin 
in different system. For an electron moving in magnetic field, the non-zero $\omega$ means the existence of magnetic field. 
For a self-propelled particle, such as bacteria or sperm, the non-zero $\omega$ represents the effective centripetal 
force generated by the interaction between beating flagella and environment.

\section{Three independent random fluctuations upon deterministic circular trajectory}

We introduce three independent variables to describe a moving circle. $\vec{R}(t)=R_{x}(t)\textbf{e}_{x}+R_{y}(t)\textbf{e}_{y}$
 represents the instantaneous center of the circle. The relative position of the particle with respect to the center 
is denoted as $\vec{r}(t)=r_{x}(t)\textbf{e}_{x}+r_{y}(t)\textbf{e}_{y}$. $r_{0}(t)$ is the relative distance between 
a particle and the center of circle. Usually ${r}_{0}(t)$ is decomposed as a constant part ${r}_{0}=const$ 
and a small random fluctuating part ${\xi}^r_{x}(t)$, i.e., ${r}_{0}(t)={r}_{0}+{\xi}^r(t)$. $\omega$ 
is the angular frequency that measures how fast the particle rotates around the center. The position of a 
circling particle is denoted as  
\begin{eqnarray}
\vec{R}(t)+\vec{r}(t)=x(t)\textbf{e}_{x}+y(t)\textbf{e}_{y}=\vec{q}(t), 
\end{eqnarray}
\begin{eqnarray}
x(t)&=&{R}_{x}(t)+r_{x}(t),\;\;\;\;r_{x}(t)={r}_{0}(t)\cos({\omega} t),\nonumber\\
y(t)&=&{R}_{y}(t)+r_{y}(t),\;\;\;\;r_{y}(t)= {r}_{0}(t)\sin({\omega} t).
\end{eqnarray}
When the circle center is drifting at much lower speed than the circling speed, the trajectory of the particle is a helical curve. In a thermal fluctuating environment, the frequent collision between the particle and other objects would push the particle away from the circular path, but the dominant deterministic force would draw it back. The stochastic deviations may be descried by a random noise force.

The random noise force has three independent sources. One is from external force which drive the circle center. 
The second comes from the tangent force which controls the rotating frequency. The third is the fluctuating centripetal force,
 it makes the radii become longer or shorter. A stochastic trajectory itself can not tell which source results in
 the ultimate trajectory. Therefore we need three independent stochastic variables.

The first random variable ${\xi}^{\omega}(t)$ is added to the deterministic angular frequency $\omega$, 
\begin{eqnarray}
x(t)&=&{R}_{x}(t)+{r}_{0}(t)\cos([{\omega}(t)+{\xi}^{\omega}(t)] t), \nonumber\\
y(t)&=&{R}_{y}(t)+{r}_{0}(t)\sin([{\omega}(t)+{\xi}^{\omega}(t)] t).
\end{eqnarray}
The second random variable ${\xi}^{r}(t)$ describes the stochastic fluctuation  of the radii of circle, i.e.,
\begin{eqnarray}
x(t)&=&{R}_{x}(t)+[{r}_{0}(t)+{\xi}^r(t)]\cos({\omega} t), \nonumber\\
y(t)&=&{R}_{y}(t)+[{r}_{0}(t)+{\xi}^r(t)]\sin({\omega} t).\nonumber\\
&&{\xi}^r(t)\ll{r}_{0}(t).
\end{eqnarray}
The amplitude of stochastic fluctuation ${\xi}^{r}(t)$ 
should be much smaller than the radii of the circle so that people are 
able to observe a fluctuating circle trajectory. If the fluctuating amplitude 
is much larger than the radii of the circle, one would just see some random 
trajectory instead of circular trajectory.

The third random fluctuation on the circle center also give us a stochastic trajectory, 
we add the random variable $\vec{\xi}^{R}={\xi}^{R}_{x}(t)+{\xi}^{R}_{y}(t)$ on the circle center,
\begin{eqnarray}
x(t)&=&[{R}_{x}(t)+{\xi}^R_{x}(t)]+{r}_{0}(t)\cos({\omega} t), \nonumber\\
y(t)&=&[{R}_{y}(t)+{\xi}^R_{y}(t)]+{r}_{0}(t)\sin({\omega} t). 
\end{eqnarray}

The three random fluctuations plays different role in different system. For example, for a charged heavy particle, 
the effective driven potential is magnetic field. Angular frequency is given by $\omega={eB}/{mc}$. The charge $e$ and mass $m$ is fixed, $c$ is the speed of light which is also a constant. Magnetic field $B$ becomes the only source of fluctuation. If one can fix magnetic field, we eliminate the fluctuation of angular frequency. The radii of the circle is also controlled by the value of magnetic field, so the fluctuation of the radii also vanished. Finally we know the stochastic trajectory is induced by the fluctuation of the circle center. The center of the charged particle is very hard to control in many body system.

For bacteria or sperm, the sources of force is versatile. Neither angular frequency, nor radii of the circle is controllable. The general way of modeling such random circling dynamics is to take into account of all the three types of fluctuation. We input the amplitude of different fluctuation from experimental observation. For instance, if the circling path is very smooth, the fluctuation of radii of the circle should be very small. If the circling velocity along the circle is almost constant, the fluctuation of angular velocity is ignorable. If the center of circle is almost static, one can ignore the fluctuation of the circle center.

\section{The calculation of probability conservation equation of single circling particle in cylindrical coordinates}

If the system has rotational symmetry, it is more convenient to use the diffusion equation expressed in polar coordinates system. For example, if one places a charge at the origin point, all the electric field vectors point out of the origin. The field has the same strength on every circle centered at the origin. Newton's equation for the circle center in polar coordinates has the same formulation as that in Cartesian coordinates, 
\begin{eqnarray}
m\frac{\partial{\vec{V}}}{\partial{t}}
=-m\vec{V}\times{\vec{\omega}}+\vec{F}-m\eta{\vec{V}}.
\end{eqnarray}
The velocity $\vec{V}=({V}_R,\;\;{V}_\theta)$ is expressed by polar coordinates,
\begin{equation}
{V}_{R}=\dot{R},\;\;{V}_\theta=R\;\dot{\theta},\;\;\;\;\;\;{V}_R\perp{V}_\theta.
\end{equation}
For the overdamped case, we set inertial term as zero, i.e., $\dot{V}_{\theta}=0$ and $\dot{V}_{R}=0$. Combining this constrain into the pair of equation of motion, 
\begin{eqnarray}
m\dot{V}_{\theta}&=&-m{\omega}{V}_{R}-m\eta{V}_{\theta}+{F}_{\theta},\nonumber\\
m\dot{V}_{R}&=&m{\omega}{V}_{\theta}-m\eta{V}_{R}+{F}_{R},\nonumber
\end{eqnarray}
we get the velocity solutions
\begin{eqnarray}\label{1polarV}
R\;\dot{\theta}&=&+\sigma_{{\theta}{\theta}}F_{{\theta}}+\sigma_{{\theta}{R}}F_{{R}},\nonumber\\
\dot{R}&=&+\sigma_{{R}{R}}F_{{R}}+\sigma_{{R}{\theta}}F_{{\theta}}.
\end{eqnarray}
Here the drifting tensor $\sigma\;\; $ are
\begin{eqnarray}
\sigma_{{\theta}{\theta}}&=&\frac{\eta}{mA^{2}},\;\;\;\;\;\sigma_{{\theta}{R}}=-\frac{\omega}{mA^{2}},\nonumber\\
\sigma_{{R}{R}}&=&\frac{\eta}{mA^{2}},\;\;\;\;\;\sigma_{{R}{\theta}}=\frac{\omega}{mA^{2}},
\end{eqnarray}
where $A^{2}=\eta^2+\omega^2$ is a normalization factor. If the circle center is situated at $(R,\theta)$, the instantaneous position of the particle in polar coordinates is given by
\begin{eqnarray}
R_c&=&R+r_0\cos(\omega{t}-\theta),\nonumber\\
\theta_c&=&\theta+\arctan\left[ \frac{r_0\sin(\omega{t}-\theta)}{R+r_0\cos(\omega{t}-\theta)}\right]. 
\end{eqnarray}
The position of the circle center is governed by Eq. (\ref{1polarV}). It is the circle center that determines the position of particle in larger scale. If we study the diffusion of the circle center, the local circular motion may be put aside. In analogy with the diffusion in Cartesian coordinates, the probability distribution function must satisfy the conservation equation,  
\begin{eqnarray}
\frac{\partial\psi}{\partial{t}}+\frac{\partial}{\partial{R}}({\psi}V_{R})
+\frac{1}{R}\frac{\partial}{\partial_{\theta}}({\psi}V_{\theta})=0.
\end{eqnarray}
For the velocity solution equation (\ref{1polarV}), the diffusion equation in polar coordinates reads, 
\begin{eqnarray}\label{1polar0}
\frac{\partial{\psi}}{\partial{t}}
&+&\frac{\partial}{\partial{R}}[\sigma_{{\theta}{\theta}}{{\psi}}F_{{\theta}}+\sigma_{{\theta}{R}}{{\psi}}F_{{R}}
] \nonumber\\
&+&\frac{1}{R}\frac{\partial}{\partial{\theta}}
[\sigma_{{R}{R}}{{\psi}}F_{{R}}+\sigma_{{R}{\theta}}{{\psi}}F_{{\theta}}]\nonumber\\
&=&0.
\end{eqnarray}
We consider a special case, the drift tensor and diffusion tensor are constant, The external driven force has radial symmetry,
\begin{equation}
F_{\theta}=0,\;\;\;F_{R}=C R. 
\end{equation} 
The diffusion equation is, 
\begin{eqnarray}\label{polar2}
\frac{\partial{\psi}}{\partial{t}}+
\sigma_{{R}{R}}C\frac{\partial{{\psi}}}{\partial{\theta}}
+\sigma_{{\theta}{R}}CR\frac{\partial{{\psi}}}{\partial{R}}
+\sigma_{{\theta}{R}}C{{\psi}}=0.
\end{eqnarray}
Even if there is no external force in $\theta$-direction, $F_{\theta}=0$, the external field in the radial direction $F_{R}$ can induce a diffusive flow in $\theta$-direction.

\section{The calculation of probability conservation equation of single circling particle in three dimensions}

Circling particle in three dimensions produces helical curve. The angular velocity of a circling particle has three components,
\begin{eqnarray}
\vec{\omega}(t)&=&{\omega}_x(t)\textbf{e}_{x}+{\omega}_y(t)\textbf{e}_{y}+{\omega}_z(t)\textbf{e}_{z},\nonumber\\
\omega^2(t)&=&\omega_x^2(t)+\omega_y^2(t)+\omega_z^2(t).
\end{eqnarray}
When the circle center is placed at the origin, the instantaneous position of the particle can be expressed by the angular velocity vector $\vec{\omega}(t)$,
\begin{eqnarray}\label{1posi3} 
{r}_{z}(t)&=&\frac{{r}_{0}}{\omega^2}[\omega_z\omega_x-\omega_x\omega_z\cos(\omega{t})-\omega_y{\omega}\sin(\omega{t})],\nonumber\\
{r}_{y}(t)&=&\frac{{r}_{0}}{\omega^2}[\omega_y\omega_x-\omega_x\omega_y\cos(\omega{t})+\omega_z{\omega}\sin(\omega{t})],\nonumber\\
{r}_{x}(t)&=&\frac{{r}_{0}}{\omega^2}[\omega_x\omega_x+(\omega_y^2+\omega_z^2)\cos(\omega{t})].
\end{eqnarray}
We place the circle center on an arbitrary curve $\vec{R}(t)$. This curve is the trajectory of the center of a circle. We keep the angular velocity vector $\vec{\omega}(t)$ parallel to the tangent vector of the $\vec{R}(t)$. The three dimensional Newton's equation of motion has the same form as the two dimensional equation, 
\begin{eqnarray}
m\frac{\partial{\vec{V}_{R}}}{\partial{t}}
=-m\vec{V}_{R}\times{\vec{\omega}}+\vec{F}-m\eta{\vec{V}_{R}}.
\end{eqnarray}
Here every vector has three components. This vector equation is composed of three coupled equations, 
\begin{eqnarray}\label{1dynamic3}
m\ddot{R}_{x}&=&-m(\dot{R}_{y}{\omega}_{z}-\dot{R}_{z}{\omega}_{y})-m\eta\dot{R}_{x}+{F}_{x},\nonumber\\
m\ddot{R}_{y}&=&-m(\dot{R}_{z}{\omega}_{x}-\dot{R}_{x}{\omega}_{z})-m\eta\dot{R}_{y}
+{F}_{y},\nonumber\\
m\ddot{R}_{z}&=&-m(\dot{R}_{x}{\omega}_{y}-\dot{R}_{y}{\omega}_{x})-m\eta\dot{R}_{z}
+{F}_{z}.
\end{eqnarray}
The velocity of the circle center in the overdamped case is the solution of Eq. (\ref{1dynamic3}) by imposing the constrains, 
$\ddot{R}_{x}=0,$ $\;\;\ddot{R}_{y}=0,$ $\;\;\ddot{R}_{z}=0,$
\begin{eqnarray}\label{3dotR} 
\dot{R}_{x}&=&{\sigma_{xx}}F_{x}+{\sigma_{xy}}F_{y}+{\sigma_{xz}}F_{z},\nonumber\\
\dot{R}_{y}&=&{\sigma_{yx}}F_{x}+{\sigma_{yy}}F_{y}+{\sigma_{yz}}F_{z},\nonumber\\
\dot{R}_{z}&=&{\sigma_{zx}}F_{x}+{\sigma_{zy}}F_{y}+{\sigma_{zz}}F_{z},
\end{eqnarray}
were the drift tensor are 
\begin{eqnarray}\label{1sigma}
\sigma_{xx}&=&\frac{1}{C_{0}^2}(\eta^2+m\omega_x^2),\;\;\;\;\;\;\;\;\;\;
\sigma_{xy}=\frac{1}{C_{0}^2}(-\eta\omega_z+\omega_x\omega_y),\nonumber\\
\sigma_{xz}&=&\frac{1}{C_{0}^2}(\eta\omega_y+\omega_x\omega_z),\;\;\;\;\;\;\;
\sigma_{yx}=\frac{1}{C_{0}^2}(\eta\omega_z+\omega_x\omega_y),\nonumber\\
\sigma_{yy}&=&\frac{1}{C_{0}^2}(\eta^2+\omega_y^2),\;\;\;\;\;\;\;\;\;\;\;\;\;
\sigma_{yz}=\frac{1}{C_{0}^2}(-\eta\omega_x+\omega_y\omega_z),\nonumber\\
\sigma_{zx}&=&\frac{1}{C_{0}^2}(-\eta\omega_y+\omega_x\omega_z),\;\;\;\;
\sigma_{zy}=\frac{1}{C_{0}^2}(\eta\omega_x+\omega_y\omega_z),\nonumber\\
\sigma_{zz}&=&\frac{1}{C_{0}^2}(\eta^2+\omega_z^2),
\end{eqnarray}
where $C_{0}^2=m\eta(\eta^2+m\omega_x^2+m\omega_y^2+m\omega_z^2)$ is the normalization factor. The drifting tensor $\sigma$ is the sum of a symmetric matrix $\sigma^{1}$ and antisymmetric matrix $\sigma^{2}$, $\sigma=\sigma^{1}+\sigma^{2},$
\begin{displaymath}
% use packages: array
\sigma^{1}=\frac{1}{C_{0}^2}\left[\begin{array}{ccc}
\eta^2+m\omega_x^2 & \omega_x\omega_y & \omega_x\omega_z \\ 
\omega_y\omega_x & \eta^2+m\omega_y^2 & \omega_y\omega_z \\ 
\omega_z\omega_x & \omega_z\omega_y & \eta^2+m\omega_z^2
\end{array}\right],\;\;\; 
\end{displaymath}
\begin{displaymath}
\sigma^{2}=\frac{1}{C_{0}^2}\left[\begin{array}{ccc}
0 & -\eta\omega_z & \eta\omega_y \\ 
\eta\omega_z & 0 & -\eta\omega_x \\ 
-\eta\omega_y & \eta\omega_x & 0
\end{array}\right].
\end{displaymath}
The equation of motion (\ref{3dotR}) governs the position of the circle center. The instantaneous position of the particle around center is determined by (\ref{1posi3}). The complete position equation of the circling particle is
\begin{eqnarray}\label{1R+r} 
{x}(t)&=&{R}_{x}+{r}_{x}(t),\nonumber\\
{y}(t)&=&{R}_{y}+{r}_{y}(t),\nonumber\\
{z}(t)&=&{R}_{z}+{r}_{z}(t).
\end{eqnarray}
This equation describes a helical curve. The three dimensions diffusion equation for a single circling particle is
\begin{equation}
\partial_t{{\psi}}+\nabla_R\cdot({{\psi}\vec{V}_{R}})=0, 
\end{equation}
where $\vec{V}_{R}=(\dot{R}_{x},\dot{R}_{y},\dot{R}_{z})$ is given by Eq. (\ref{3dotR}).

If the friction coefficient $\eta$, drift tensor ${\sigma}_{ij}$ are constants, 
the explicit expansion of the diffusion equation is
\begin{eqnarray}
\frac{\partial{\psi}}{\partial{t}}
&+&[{\sigma_{xx}}F_{x}+{\sigma_{xy}}F_{y}+{\sigma_{xz}}F_{z}]\;\partial_{R_x}{{{\psi}}}\nonumber\\
&+&[{\sigma_{yx}}F_{x}+{\sigma_{yy}}F_{y}+{\sigma_{yz}}F_{z}]\;\partial_{R_y}{{{\psi}}}\nonumber\\
&+&[{\sigma_{zx}}F_{x}+{\sigma_{zy}}F_{y}+{\sigma_{zz}}F_{z}]\;\partial_{R_z}{{{\psi}}}\nonumber\\
&+&[{\sigma_{xx}}\partial_{R_x}F_{x}+{\sigma_{xy}}\partial_{R_x}F_{y}+{\sigma_{xz}}\partial_{R_x}F_{z}]\;{{{\psi}}}\nonumber\\
&+&[{\sigma_{yx}}\partial_{R_y}F_{x}+{\sigma_{yy}}\partial_{R_y}F_{y}+{\sigma_{yz}}\partial_{R_y}F_{z}]\;{{{\psi}}}\nonumber\\
&+&[{\sigma_{zx}}\partial_{R_z}F_{x}+{\sigma_{zy}}\partial_{R_z}F_{y}+{\sigma_{zz}}\partial_{R_z}F_{z}]\;{{{\psi}}}\nonumber\\
&=&0.
\end{eqnarray}
The components of the drifting tensor are not independent. As long as the angular velocity is not zero, the external force in $X$-direction $F_{x}$ will induce the motion in $Y-$ and $Z-$direction, so does the other force component, $F_{y}$ and $F_{z}$. This is the special dynamics of circling particles.

{\section{The derivation of density evolution equation of many circling particles with cylindrical density gradient}}

If the particles are confined in a small disk around the origin point in the beginning, 
the density gradient generated during diffusion has SO(2) rotational symmetry. 
We choose polar coordinates to express the diffusion equation. Newton's equation of motion is 
\begin{equation}
mN_{s}\frac{\partial{\vec{V}}}{\partial{t}}
=-mN_{s}\vec{V}\times{\vec{\omega}}+N_{s}\vec{F}-D\nabla{N_{s}}-mN_{s}\eta{\vec{V}}.
\end{equation}
The explicit formulation of this vector equation reads, 
\begin{eqnarray}
m\dot{V}_{\theta}&=&-m{\omega}{V}_{R}-\frac{D_{\theta}}{R}\frac{\partial_{\theta}{N_{s}}}{N_{s}}-m\eta{V}_{\theta}+{F}_{\theta},\nonumber\\
m\dot{V}_{R}&=&m{\omega}{V}_{\theta}-D_{R}\frac{\partial_{R}{N_{s}}}{N_{s}}-m\eta{V}_{R}+{F}_{R},\nonumber
\end{eqnarray}
$D_{\theta}$ is the diffusion coefficient tangent to the circle. $D_{R}$ is the diffusion coefficient along the radius. In the overdamped case, $m\dot{V}_{\theta}=0$ and $m\dot{V}_{R}=0$. We solve the equations above to get explicit representation of velocity, 
\begin{eqnarray}\label{polarV}
R\;\dot{\theta}&=&-{D_{{\theta}{R}}}\frac{\partial_{{R}}{N_{s}}}{N_{s}}-
{D_{{\theta}{\theta}}}\frac{1}{R}\frac{\partial_{{\theta}}{N_{s}}}{N_{s}}+\sigma_{{\theta}{\theta}}F_{{\theta}}+\sigma_{{\theta}{R}}F_{{R}},\nonumber\\
\dot{R}&=&-{D_{{R}{\theta}}}\frac{1}{R}\frac{\partial_{{\theta}}{N_{s}}}{N_{s}}-
{D_{{R}{R}}}\frac{\partial_{{R}}{N_{s}}}{N_{s}}+\sigma_{{R}{R}}F_{{R}}+\sigma_{{R}{\theta}}F_{{\theta}}.\nonumber\\
\end{eqnarray}
Here the diffusion tensor $D$ and drifting tensor $\sigma\;\; $ are
\begin{eqnarray}
D_{{\theta}{\theta}}&=&\frac{{D}_{{\theta}}\eta}{mA^{2}},\;\;D_{{\theta}{R}}=-\frac{D_{{R}}\omega}{mA^{2}},\nonumber\\
D_{{R}{R}}&=&\frac{{D}_{{R}}\eta}{mA^{2}},\;\;D_{{R}{\theta}}=\frac{D_{{\theta}}\omega}{mA^{2}},\nonumber\\
\sigma_{{\theta}{\theta}}&=&\frac{\eta}{mA^{2}},\;\;\;\;\;\sigma_{{\theta}{R}}=-\frac{\omega}{mA^{2}},\nonumber\\
\sigma_{{R}{R}}&=&\frac{\eta}{mA^{2}},\;\;\;\;\;\sigma_{{R}{\theta}}=\frac{\omega}{mA^{2}},
\end{eqnarray}
where $A^{2}=\eta^2+\omega^2$ is a normalization factor. The transverse flow induced by velocity gradient is given by
\begin{eqnarray}
J_{R}&=&\frac{\omega}{R}\partial_{\theta}V_{R}=-\frac{\omega}{R}\partial_{\theta}({D_{{R}{\theta}}}\frac{1}{R}\frac{\partial_{{\theta}}{N_{s}}}{N_{s}}+{D_{{R}{R}}}\frac{\partial_{{R}}{N_{s}}}{N_{s}})\nonumber\\
&+&\frac{\omega}{R}\partial_{\theta}(\sigma_{{R}{R}}F_{{R}}+\sigma_{{R}{\theta}}F_{{\theta}}),\nonumber\\
J_{\theta}&=&-\omega\partial_{R}V_{\theta}=\omega\partial_{R}({D_{{\theta}{R}}}\frac{\partial_{{R}}{N_{s}}}{N_{s}}+
{D_{{\theta}{\theta}}}\frac{1}{R}\frac{\partial_{{\theta}}{N_{s}}}{N_{s}})\nonumber\\
&-&\omega\partial_{R}(\sigma_{{R}{R}}F_{{R}}+\sigma_{{R}{\theta}}F_{\theta}).
\end{eqnarray}
The derivative of velocity is acceleration. The flow field $\vec{J}$ is actually the off-diagonal acceleration. Substituting the velocity equation (\ref{polarV}) into the conservation equation,
\begin{eqnarray}
\frac{\partial{N}_{s}}{\partial{t}}+\frac{\partial}{\partial{R}}({N}_{s}V_{R})
+\frac{1}{R}\frac{\partial}{\partial_{\theta}}({N}_{s}V_{\theta})=0,
\end{eqnarray}
we get the diffusion equation in polar coordinates, 
\begin{eqnarray}\label{polar0}
\frac{\partial{N}_{s}}{\partial{t}}
&-&\frac{\partial}{\partial{R}}
[{D_{{R}{\theta}}}\frac{1}{R}\frac{\partial}{\partial{\theta}}{{N_{s}}}+{D_{{R}{R}}}{\frac{\partial}{\partial{R}}{N_{s}}}]\nonumber\\
&+&\frac{\partial}{\partial{R}}[\sigma_{{\theta}{\theta}}{N_{s}}F_{{\theta}}+\sigma_{{\theta}{R}}{N_{s}}F_{{R}}
] \nonumber\\
&-&\frac{1}{R}\frac{\partial}{\partial{\theta}}
[{D_{{\theta}{R}}}{\frac{\partial}{\partial{R}}{N_{s}}}+
{D_{{\theta}{\theta}}}\frac{1}{R}\frac{\partial}{\partial{\theta}}{{N_{s}}}] \nonumber\\
&+&\frac{1}{R}\frac{\partial}{\partial{\theta}}
[\sigma_{{R}{R}}{N_{s}}F_{{R}}+\sigma_{{R}{\theta}}{N_{s}}F_{{\theta}}]\nonumber\\
&=&0.
\end{eqnarray}

{\section{The derivation of three dimensional density evolution equation of many circling particles}}

\begin{figure}[htbp]
\centering
\par
\hspace{-0.9cm}
\begin{center}\hspace{-0.2cm}
\includegraphics[width=0.47\textwidth]{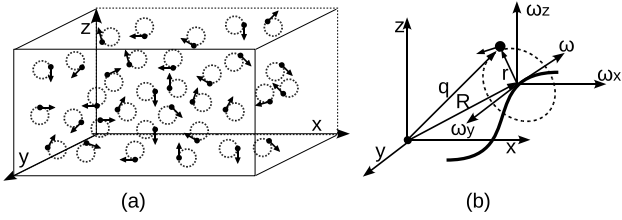}
\end{center}
\caption{\label{angular} (a) The circling particles swimming in three dimensions. (b) The decomposition of the motion of a circling particle in three dimensions. The local circling motion is around $\omega$. The motion of the circle's center is along the dark thick curve. $\omega$ is parallel to the tangent vector of the center line(the dark thick curve). } \vspace{-0.2cm}
\end{figure}

Circling particles moving in three dimensional space draws a helical trajectory. A classical electron or ion moving in three dimensions follows a helical trajectory. The center line of the helical path is magnetic field line. We will study a box of circling particles in three dimensions(Fig. \ref{angular}). Following the same process of modeling the diffusion phenomena in two dimensions, we took the unit scale as the radius of the circle. The Newtonian equation of motion a unit box of $N_s$ particles reads
\begin{equation}
mN_{s}\frac{\partial{\vec{V}}}{\partial{t}}
=-mN_{s}\vec{V}\times{\vec{\omega}}+N_{s}\vec{F}-D\nabla{N_{s}}-mN_{s}\eta{\vec{V}}.
\end{equation}
Here every vector has three components. The equation of motion is composed of three coupled equations, 
\begin{eqnarray}\label{dynamic3}
m\ddot{R}_{x}&=&-m(\dot{R}_{y}{\omega}_{z}-\dot{R}_{z}{\omega}_{y})-
D_x\frac{\partial_{Rx}{N_{s}}}{N_{s}}-m\eta\dot{R}_{x}+{F}_{x},\nonumber\\
m\ddot{R}_{y}&=&-m(\dot{R}_{z}{\omega}_{x}-\dot{R}_{x}{\omega}_{z})-D_y\frac{\partial_{Ry}{N_{s}}}{N_{s}}-m\eta\dot{R}_{y}
+{F}_{y},\nonumber\\
m\ddot{R}_{z}&=&-m(\dot{R}_{x}{\omega}_{y}-\dot{R}_{y}{\omega}_{x})-D_z\frac{\partial_{Rz}{N_{s}}}{N_{s}}-m\eta\dot{R}_{z}
+{F}_{z}.\nonumber
\end{eqnarray}
We study the overdamped case, $\ddot{R}_{x}=0,$ $\;\;\ddot{R}_{y}=0,$ $\;\;\ddot{R}_{z}=0.$. The velocity of the circle center is
\begin{eqnarray}\label{dotR} 
\dot{R}_{x}&=&-{D_{xx}}\frac{\partial_{Rx}{N_{s}}}{N_{s}}-{D_{xy}}\frac{\partial_{Ry}{N_{s}}}{N_{s}}-{D_{xz}}\frac{\partial_{Rz}{N_{s}}}{N_{s}}\nonumber\\
&+&{\sigma_{xx}}F_{x}+{\sigma_{xy}}F_{y}+{\sigma_{xz}}F_{z},\nonumber\\
\dot{R}_{y}&=&-{D_{yx}}\frac{\partial_{Rx}{N_{s}}}{N_{s}}-{D_{yy}}\frac{\partial_{Rx}{N_{s}}}{N_{s}}-{D_{yz}}\frac{\partial_{Rx}{N_{s}}}{N_{s}},\nonumber\\
&+&{\sigma_{yx}}F_{x}+{\sigma_{yy}}F_{y}+{\sigma_{yz}}F_{z},\nonumber\\
\dot{R}_{z}&=&-{D_{zx}}\frac{\partial_{Rx}{N_{s}}}{N_{s}}-{D_{zy}}\frac{\partial_{Rx}{N_{s}}}{N_{s}}-{D_{zz}}\frac{\partial_{Rx}{N_{s}}}{N_{s}}\nonumber\\
&+&{\sigma_{zx}}F_{x}+{\sigma_{zy}}F_{y}+{\sigma_{zz}}F_{z}.
\end{eqnarray}
Both the diffusion tensor and drift tensor are $3\times3$ matrix, the relation between diffusion tensor and drifting tensor are 
\begin{eqnarray}
D_{xx}&=&D_{x}\sigma_{xx},\;\;\;\;  
D_{xy}=D_{y}\sigma_{xy},\;\;\;\;  
D_{xz}=D_{z}\sigma_{xz},\;\;\;\;  \nonumber\\
D_{yx}&=&D_{x}\sigma_{yx},\;\;\;\;  
D_{yy}=D_{y}\sigma_{yy},\;\;\;\;  
D_{yz}=D_{z}\sigma_{yz},\;\;\;\;  \nonumber\\
D_{zx}&=&D_{x}\sigma_{zx},\;\;\;\;  
D_{zy}=D_{y}\sigma_{zy},\;\;\;\;  
D_{zz}=D_{z}\sigma_{zz},\;\;\;\; 
\end{eqnarray}
The drifting tensor is defined by Eq. (\ref{1sigma}). 
Following the same calculation for two dimensional diffusion theory, 
we get the diffusion equation in three dimensions by substituting the velocity 
in to the conservation equation $\partial_t{{N}_{s}}+\nabla_R\cdot{{N}_{s}\vec{V}}=0$. 
If the friction coefficient $\eta=\eta(R_x,R_y,R_z)$, drift tensor ${\sigma}_{ij}=\sigma(R_x,R_y,R_z)$ 
and diffusion tensor $D_{i}=D_{i}(R_x,R_y,R_z)$ are all state-dependent, the three dimensional diffusion equation is
\begin{equation}\label{difu3D0}
\frac{\partial{N}_{s}}{\partial{t}}+\sum_{ij}\frac{\partial}{{\partial{R}}_i}[{\sigma}_{ij}{F}_{j}N_{s}]
-\sum_{ij}\frac{\partial}{{\partial{R}}_i}[{D}_{ij}\frac{\partial}{{\partial{R}}_j}{N}_{s}]=0.
\end{equation}
For the special case that the friction coefficient $\eta$, drift tensor ${\sigma}_{ij}$ and diffusion tensor $D_{i}$ are constants, we get an explicit expansion of the diffusion equation,
\begin{eqnarray}
\frac{\partial{N}_{s}}{\partial{t}}
&-&[D_{xy}+D_{yx}]\partial_{R_x}\partial_{R_y}N_{s}-D_{xx}\partial_{R_x}^{2}N_{s}\nonumber\\
&-&[D_{xz}+D_{zx}]\partial_{R_x}\partial_{R_z}N_{s}-D_{yy}\partial_{R_y}^{2}N_{s}\nonumber\\
&-&[D_{zy}+D_{yz}]\partial_{R_y}\partial_{R_z}N_{s}-D_{zz}\partial_{R_z}^{2}N_{s}\nonumber\\
&+&[{\sigma_{xx}}F_{x}+{\sigma_{xy}}F_{y}+{\sigma_{xz}}F_{z}]\;\partial_{R_x}{{N_{s}}}\nonumber\\
&+&[{\sigma_{yx}}F_{x}+{\sigma_{yy}}F_{y}+{\sigma_{yz}}F_{z}]\;\partial_{R_y}{{N_{s}}}\nonumber\\
&+&[{\sigma_{zx}}F_{x}+{\sigma_{zy}}F_{y}+{\sigma_{zz}}F_{z}]\;\partial_{R_z}{{N_{s}}}\nonumber\\
&+&[{\sigma_{xx}}\partial_{R_x}F_{x}+{\sigma_{xy}}\partial_{R_x}F_{y}+{\sigma_{xz}}\partial_{R_x}F_{z}]\;{{N_{s}}}\nonumber\\
&+&[{\sigma_{yx}}\partial_{R_y}F_{x}+{\sigma_{yy}}\partial_{R_y}F_{y}+{\sigma_{yz}}\partial_{R_y}F_{z}]\;{{N_{s}}}\nonumber\\
&+&[{\sigma_{zx}}\partial_{R_z}F_{x}+{\sigma_{zy}}\partial_{R_z}F_{y}+{\sigma_{zz}}\partial_{R_z}F_{z}]\;{{N_{s}}}\nonumber\\
&=&0.
\end{eqnarray}
One can choose different coordinates system to reformulate the three dimensional diffusion equation for the convenience of physics, especially when the physical system has rotational symmetry. If the initial distribution of particle has spherical symmetry, 
it is better to express the diffusion equation by three dimensional polar coordinates. When the density gradient distribution expands away from the center, there would induce some transverse flow distributed on the spherical surface. There is a topological constrain on the distribution of the flows on a sphere, it states there must exist two singular points at which the flows contradict one another. The flow on the sphere tend to fuse into the flow along radius.

\end{document}